\documentstyle[pra,aps,epsf,twocolumn]{revtex}

\newcommand{\boldvec}[1]{\mbox{\boldmath$#1$}}
\newcommand{\smallvec}[1]{\mbox{\boldmath$\scriptstyle#1$}}
\newcommand{\annhilate}[2]{\hat{#1}^{\phantom\dag}_{{#2}}}
\newcommand{\create}[2]{\hat{#1}^{\dag}_{{#2}}}

\newcommand{\be}{{\bf 1}}
\newcommand{\bz}{{\bf 2}}
\newcommand{\bd}{{\bf 3}}
\newcommand{\bv}{{\bf 4}}

\begin{document}

\draft

\title{Resonance Superfluidity: Renormalization of Resonance
  Scattering Theory}

\author{S.J.J.M.F. Kokkelmans$^1$, J.N. Milstein$^1$, M.L.
  Chiofalo$^2$, R.  Walser$^1$, and M.J. Holland$^1$}

\address{$^1$JILA, University of Colorado and National Institute of
  Standards and Technology, Boulder, Colorado 80309-0440, U.S.A.}

\address{$^2$INFM and Classe di Scienze, Scuola Normale Superiore,
  Piazza dei Cavalieri 7, I-56126, Pisa, Italy}

\date{\today}

\wideabs{

\maketitle

\begin{abstract}
  We derive a theory of superfluidity for a dilute Fermi gas that is
  valid when scattering resonances are present. The treatment of a
  resonance in many-body atomic physics requires a novel mean-field
  approach starting from an unconventional microscopic Hamiltonian.
  The mean-field equations incorporate the microscopic scattering
  physics, and the solutions to these equations reproduce the
  energy-dependent scattering properties. This theory describes the
  high-$T_c$ behavior of the system, and predicts a
  value of $T_c$ which is a significant fraction of the Fermi
  temperature. It is shown that this novel mean-field approach does
  not break down for typical experimental circumstances, even at
  detunings close to resonance. As an example of the application of
  our theory we investigate the feasibility for achieving
  superfluidity in an ultracold gas of fermionic $^6$Li.
\end{abstract}

}

\section{Introduction}

The remarkable accomplishment of reaching the regime of quantum
degeneracy~\cite{bec} in a variety of ultracold atomic gases enabled the
examination of superfluid phenomena in a diverse range of
novel quantum systems. Already many elementary aspects of superfluid
phenomena have been observed in bosonic systems including
vortices~\cite{vortices}. The challenge of achieving
superfluidity in a Fermi gas remains, however, although it appears
possible that this situation may change in the near future. A number
of candidate systems for realizing superfluidity in a fermionic gas
appear very promising and it is currently the goal of several
experimental efforts to get into the required regime to observe the
superfluid phase transition.  So far both fermionic
potassium~\cite{dfg} and lithium~\cite{truscott,schreck} have been
cooled to the microkelvin regime and are well below the Fermi
temperature by now---a precursor step for superfluidity.

In order to make the superfluid phase transition experimentally
accessible, it will likely be necessary to utilize the rich internal
hyperfine structure of atomic collisions. Scattering resonances, in
particular, may prove to be extremely important since they potentially
allow a significant enhancement of the strength of the atomic
interactions. It is anticipated that by utilizing such a scattering
resonance one may dramatically increase the critical temperature at
which the system becomes unstable towards the formation of Cooper
pairs, thus bringing the critical temperature into the experimentally
accessible regime.

In spite of its promise, this situation poses a number of fundamental
theoretical problems which must be addressed in order to provide an
adequate minimal description of the critical behavior. The scope of
the complexities that arise in treating a scattering resonance can be
seen by examining the convergence of the quantum kinetic perturbation
theory of the dilute gas. In this theory the small parameter is known
as the gaseous parameter; defined as $\sqrt{na^3}$ where $n$ is the particle
density and $a$ is the scattering length. Formally, when the
scattering length is increased to the value at which $na^3\approx1$,
conventional perturbation theory breaks down\cite{beliaev,abrikosov}.
This situation is commonly associated with the theoretical treatment
of strongly interacting fermionic systems where high-order
correlations must be treated explicitly.

In this paper, we show that an unconventional mean-field theory can
still be appropriately exploited under the condition that the
characteristic range $R$ of the potential is such that $nR^3\ll1$
(while $na^3\gtrsim1$). The core issue is that around a resonance, the
cross-section becomes strongly dependent on the scattering energy.
This occurs when either a bound state lies just below threshold, or
when a quasi-bound state lies just above the edge of the collision
continuum.  In both cases, the scattering length---evaluated by
considering the zero energy limit of the scattering phase shift---does
not characterize the full scattering physics over the complete energy
range of interest, even when in practice this may cover a range of
only a few microkelvin.

The paper is outlined as follows. In Section II, we present a
systematic derivation of the renormalized potentials for an effective
many-body Hamiltonian. This requires a detailed analysis of
coupled-channels scattering. In Section III, we derive the resonance
mean-field theory. In Section IV, we present the thermodynamic
solutions allowing for resonance superfluidity. We apply our theory to
the specific case of $^6$Li and determine the critical temperature for
the superfluid phase transition. In Section V, we consider the validity
of the mean field approach in the case of resonance coupling, and
establish the equivalence with previous diagrammatic calculations of
the crossover regime between fermionic and bosonic superconductivity.

\section{Two-Body Resonance Scattering} \label{twobodyresscat}

\begin{figure}
\begin{center}\
  \epsfysize=60mm \epsfbox{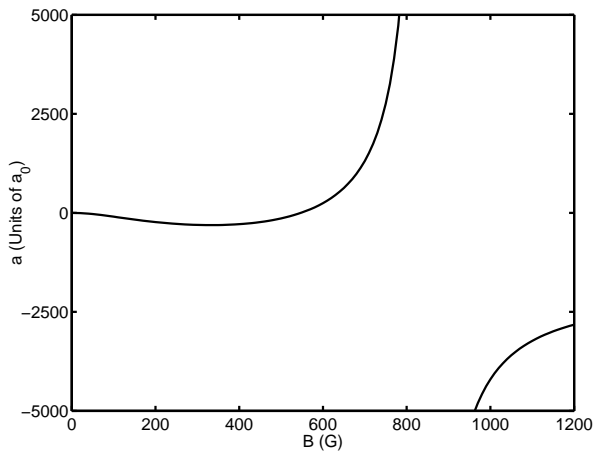}
\end{center}
\caption{Scattering length as a function of magnetic field, for the
  $(f,m_f)=(1/2,-1/2)$ and $(1/2,1/2)$ mixed spin channel of $^6$Li.}
\label{fig1}
\end{figure}

The position of the last bound state in the interatomic interaction
potentials generally has a crucial effect on the scattering
properties. In a single-channel system, the scattering process becomes
resonant when a bound state is close to threshold. In a multi-channel
system the incoming channel (which is always open) may be coupled
during the collision to other open or closed channels corresponding to
different spin configurations. When a bound state in a closed channel
lies near the zero of the collision energy continuum, a Feshbach
resonance~\cite{feshbach} may occur, giving rise to scattering
properties which are tunable by an external magnetic field. The tuning
dependence arises from the magnetic moment difference $\Delta\mu^{\rm
  mag}$ between the open and closed channels~\cite{tiesinga}.  This
gives rise to a characteristic dispersive behavior of the $s$-wave
scattering length at fields close to resonance given by
\begin{equation}
  a = a_{{\rm bg}}\left( 1-\frac{\Delta B}{B-B_0} \right),
  \label{reala}
\end{equation}
where $a_{\rm bg}$ is the background value which may itself depend
weakly on magnetic field. The field-width of the resonance is given by
$\Delta B$, and the bound state crosses threshold at a field-value
$B_0$. The field-detuning can be converted into an energy-detuning
$\bar\nu$ by the relation $\bar\nu=(B-B_0)\Delta\mu^{\rm mag}$. An
example of such a resonance is given in Fig.~\ref{fig1}, where a
coupled channels calculation is shown of the scattering length of
$^6$Li for collisions between atoms in the $(f,m_f)=(1/2,-1/2)$ and
$(1/2,1/2)$ state~\cite{servaaspriv}. The background scattering length
changes slowly as a function of magnetic field due to a
field-dependent mixing of a second resonance which comes from the
triplet potential. This full coupled channels calculation includes the
state-of-the-art interatomic potentials~\cite{abeelen} and the
complete internal hyperfine structure~\cite{stoof}.

The scattering length is often used in many-body theory to describe
interactions in the $s$-wave regime. That the scattering length
completely encapsulates the collision physics over relevant energy
scales is implicitly assumed in the derivation of the conventional
Bardeen-Cooper-Schrieffer (BCS) theory for degenerate
gases~\cite{bcs,leggett}, as well as the Gross-Pitaevskii description
of Bose-Einstein condensates.  However, the scattering length is only
a useful concept in the energy regime where the $s$-wave scattering
phase shift $\delta_{0}$ depends linearly on the wavenumber $k$, i.e.\
$\delta_{0}=-ka$. For a Feshbach resonance system at a finite
temperature there will always be a magnetic field value where this
approximation breaks down and the scattering properties become
strongly energy dependent.  In close proximity to a resonance, the
scattering process then has to be treated by means of the energy
dependent $T$-matrix.

Only the exact interatomic interaction will reproduce the full
$T$-matrix over all energy scales. However, since only collision
energies in the ultracold regime (of order microkelvin) are relevant,
a much simpler description is possible. If the scattering length does
not completely characterize the low energy scattering behavior in the
presence of a resonance, what is the minimal set of parameters which
do?

\begin{figure}[h]
\begin{center}\
  \epsfysize=80mm \epsfbox{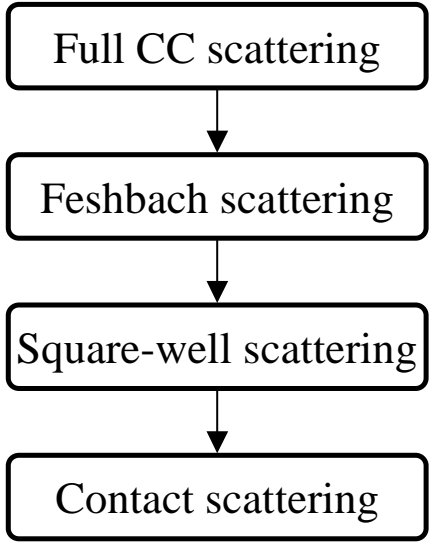}
\end{center}
\caption{
  Sequence of theoretical steps involved in formulating a renormalized
  scattering model of resonance physics for low energy scattering. The
  starting point is a full coupled channels (CC) calculation, which
  leads via an equivalent Feshbach theory, and an analytic coupled square
  well theory, to a contact potential scattering theory which gives the
  renormalized equations for the resonance system.}
\label{fig2}
\end{figure}

As illustrated in Fig.~\ref{fig2}, we proceed to systematically
resolve this question by the following steps. We start from a
numerical solution of the complete coupled channels scattering problem
for a given real physical system. In Section~\ref{FBrep} we
demonstrate that the results of these full numerical calculations can
be adequately replicated by giving an analytic description of
resonance scattering provided by Feshbach's resonance theory. The
point of this connection is to demonstrate that only a few parameters
are necessary to account for all the collision properties. This
implies that the scattering model is not unique. There are many
microscopic models which could be described by the same Feshbach
theory. In Section~\ref{doublewell} we show this explicitly by
presenting a simple double well model for which analytic solutions are
accessible.  Thereby we derive a limiting model in which the range of
the square well potentials and coupling matrix elements are taken to
zero. This leads in Section~\ref{cutoffscat} to a scattering model of
contact potentials. We show that such a scattering solution is able to
reproduce well the results of the intricate full numerical model we
began with. The utility of this result is that, as will be apparent
later, it greatly simplifies the many-body theoretic description.

\subsection{Feshbach resonance theory} \label{FBrep}

Here we briefly describe the Feshbach resonance formalism and derive
the elastic $S$- and $T$-matrices for two-body scattering. These
matrices are related to the transition probabilities for scattering
from an initial channel $\alpha$ to a final channel $\beta$. A more
detailed treatment of this formalism can be found in the
literature~\cite{feshbach}.

In Feshbach resonance theory two projection operators $P$ and $Q$ are
introduced which project onto the subspaces ${\cal P}$ and ${\cal Q}$.
These subspaces are two orthogonal components which together span the
full Hilbert space of both scattering and bound wavefunctions.  The
open and closed channels are contained in $\cal P$ and $\cal Q$,
respectively. The operators $P$ and $Q$ split the Schr\"odinger
equation for the two-body problem into two parts:

\begin{eqnarray}
  (E-H_{PP})|\psi^P\rangle &=&H_{PQ}|\psi^Q\rangle ,  \label{Peqn} \\
  (E-H_{QQ})|\psi^Q\rangle &=&H_{QP}|\psi^P\rangle , \label{Qeqn}
\end{eqnarray}
where $H_{PP}=PHP$, $H_{PQ}=PHQ$, etc., and $\psi$ is the total
scattering wavefunction. The projections on the two sub-spaces are
indicated by $P|\psi\rangle=|\psi^P\rangle$ and
$Q|\psi\rangle=|\psi^Q\rangle$. The Hamiltonian $H=H_0+V$ consists
of the sum of the single-particle interactions $H_0$ and the two-body
interaction $V$. Eq.~(\ref{Qeqn}) can be formally solved

\begin{equation}
  |\psi^Q\rangle=\frac{1}{E^+-H_{QQ}}H_{QP} |\psi^P\rangle,
\end{equation}
where $E^+=E+i\delta$ with $\delta$ approaching zero from positive
values.  Substituting this result into Eq.~(\ref{Peqn}), the open channels
equation can be written as $(E-H_{ {\rm eff}})|\psi^P\rangle=0$, where

\begin{equation}
  H_{{\rm eff}}=H_{PP}+H_{PQ}\frac{1}{E^+-H_{QQ}}H_{QP}.
\end{equation}
The resolvant operator is now expanded in the discrete and continuum
eigenstates of $ H_{QQ}$:

\begin{eqnarray}  \label{expansion}
  H_{{\rm eff}}=H_{PP}&+&\sum_i \frac{H_{PQ}|\phi_i\rangle
    \langle\phi_i| H_{QP}}{E-\epsilon_i} \\
  &+&\int \frac{H_{PQ}|\phi(\epsilon)\rangle\langle\phi(\epsilon)|
    H_{QP}}{E^+-\epsilon}d\epsilon. \nonumber
\end{eqnarray}
Here the $\epsilon_i$'s are the uncoupled bound-state eigenvalues.  In
practice, only a few bound states will significantly affect the
open-channel properties. In this paper, we will consider either one or
two bound states and neglect the continuum expansion in
Eq.~(\ref{expansion}). Then the formal solution for $|\psi^P\rangle$
is given by

\begin{equation}
  |\psi^P\rangle =|\psi_{\alpha}^{P+}\rangle
  +\frac{1}{E^{+}-H_{PP}}\sum_{i}\frac{H_{PQ}|\phi
_{i}\rangle \langle \phi _{i}|H_{QP}|\psi^P \rangle }{E-\epsilon_{i}},
\label{openchansolu}
\end{equation}
where $|\psi_{\alpha}^{P+}\rangle$ is the eigenstate of the direct
interaction $H_{PP}$ that satisfies the outgoing wave boundary
condition in channel $\alpha$. By multiplying from the left with
$\langle \chi_\beta|V$, where $|\chi_\beta\rangle$ is an unscattered
state in the outgoing channel $\beta$, the left-hand-side becomes the
$\cal T$-matrix for the total scattering process.  The unscattered
state is related to the scattering wave-function
$|\psi_\beta^{P-}\rangle$ with incoming boundary conditions via

\begin{equation}
  |\psi_\beta^{P-}\rangle=|\chi_\beta\rangle
  +\frac{V}{E^--H_{PP}}|\chi_\beta\rangle.
\end{equation}
The $\cal T$-matrix giving the transition amplitude is then

\begin{equation}
  {\cal T}_{\beta\alpha}={\cal T}_{\beta\alpha}^P +
  \sum_{i}\frac{\langle \psi_\beta^{P-}| H_{PQ}|\phi_{i}\rangle
    \langle \phi _{i}|H_{QP}|\psi^P \rangle }{E-\epsilon_{i}},
\label{FBtmatrix}
\end{equation}
where ${\cal T}_{\beta\alpha}^P$ is the amplitude for the direct
(non-resonant) process. From the $\cal T$-matrix we can easily go to
the $S$-matrix that is defined as
$S_{\beta\alpha}=\langle\psi_{\beta}^-|\psi_\alpha^+\rangle$. Since we
consider $s$-wave scattering only, in our case there exists a simple
relation between the $S$- and $\cal T$-matrix: $S_{\beta\alpha}=1-2\pi
i {\cal T}_{\beta\alpha}$ \cite{taylor}, and this allows us to rewrite
Eq.~(\ref{FBtmatrix}) as

\begin{equation}
  S_{\beta\alpha}=S_{\beta\alpha}^P-\sum_\gamma S_{\beta\gamma}^P
  \sum_i\frac{2\pi i \langle \psi_\gamma^+| H_{PQ}|\phi_{i}\rangle
    \langle \phi _{i}|H_{QP}|\psi^P \rangle }{E-\epsilon_{i}}.
\label{generalFBsmat}
\end{equation}
The non-resonant factors $S^{P}_{\beta\gamma}$ describe the direct
scattering process from an open channel $\gamma$ to the outgoing
channel $\beta$. Returning to Eq.~(\ref{openchansolu}), we can solve
for the component $\langle\phi_i|H_{QP}|\psi^P\rangle$ by multiplying both sides
with $\langle\phi_i|H_{QP}$.

\subsubsection{Single resonance}
For the case of only one resonant bound state and only one open
channel, the solution of Eq.~(\ref{openchansolu}) gives rise to the following
elastic $S$-matrix element (we will omit now the incoming channel label
$\alpha$):

\begin{equation}
  S=S^{P}\left[ 1-\frac{2\pi i|\langle \psi ^{P+}|H_{PQ}|\phi
      _{1}\rangle |^{2} }{E-\epsilon _{1}-\langle \phi
      _{1}|H_{QP}\frac{1}{E^{+}-H_{PP}} H_{PQ}|\phi _{1}\rangle
      }\right].
\end{equation}
The non-resonant $S$-matrix is related to the background scattering
length via $S^P=\exp[-2ika_{\rm bg}]$.  The term in the numerator
gives rise to the energy-width of the resonance, $\Gamma =2\pi
|\langle \psi ^{P+}|H_{PQ}|\phi _{1}\rangle |^{2}$, which is
proportional to the incoming wavenumber $k$ and coupling constant
$\bar g_1$ \cite{moerdijk}. The bracket in the denominator gives rise
to a shift of the bound-state energy, and to an additional width term
$i\Gamma/2$.  When we denote the energy-shift between the collision
continuum and the bound state by $\bar\nu_1$, and
represent the kinetic energy simply by $\hbar ^{2}k^{2}/m$, the
$S$-matrix element can be rewritten as

\begin{equation}
  S(k)=e^{-2ika_{\rm bg}}\left[ 1-\frac{2ik|\bar g_1|^{2}}{-\frac{4\pi
        \hbar ^{2} }{m}(\bar\nu_1-\frac{\hbar ^{2}k^{2}}{m }) +ik|\bar
      g_1|^{2}}\right] .
\end{equation}
The resulting total scattering length has exactly the dispersive
lineshape for the resonant scattering length which we presented
originally as Eq.~(\ref{reala}).

\subsubsection{Double resonance}
Often more than one resonance may need to be considered. For example,
the scattering properties for the $(1/2,-1/2) +(1/2,1/2)$ channel of
$^6$Li are dominated by a combination of two resonances: a triplet
potential resonance and a Feshbach resonance. This can be clearly seen
from Fig.~\ref{fig1}, where the residual scattering length, which
would arise in the absence of the Feshbach resonance coupling, would
be very large and negative and vary with magnetic field. This can be
compared with the value of the non-resonant background scattering
length for the triplet potential for Li which is only $31\,a_0$, which
is an accurate measure of the characteristic range of this potential.
An adequate scattering model for this system therefore requires
inclusion of both bound-state resonances. Since for $^6$Li the
coupling between these two bound states is small, it will be neglected
in the double resonance model presented here. The double-resonance
$S$-matrix, with again only one open channel, follows then from
Eq.~(\ref{generalFBsmat}) and includes a summation over two bound
states. After solving for the two components
$\langle\phi_i|H_{QP}|\psi^P\rangle$ of wave function $|\psi^P\rangle$, the
$S$-matrix can be written as

\begin{equation}
  S(k) =e^{-2ika_{\rm bg}}\left[ 1- \frac{2ik(|\bar
      g_{1}|^{2}\Delta_2+|\bar g_{2}|^{2}\Delta_1)} {ik(|\bar
      g_{1}|^{2}\Delta_2+|\bar g_{2}|^{2}\Delta_1)-\Delta_1\Delta_2 }
  \right] . \label{doubleressmat}
\end{equation}
with $\Delta_1=(\bar \nu_1-\hbar^{2}k^{2}/m)4\pi\hbar^{2}/m$, where
$\bar \nu_1$ and $\bar g_1$ are the detuning and coupling strengths
for state 1.  Equivalent definitions are used for state 2. Later we
will show that this simple analytic Feshbach scattering model mimics
the coupled channels calculation of $^6$Li. The parameters of this
model, which are related to the positions and widths of the last bound
states, can be directly found from a plot of the scattering length
versus magnetic field as given, for example, by Fig.~\ref{fig1}. The
scattering length behavior should be reproduced by the analytic
expression for the scattering length following from
Eq.~(\ref{doubleressmat}):

\begin{equation}
  a=a_{bg}-\frac{m}{4\pi\hbar^2}\left( \frac{|\bar g_1|^2}{\bar \nu_1}
    +\frac{|\bar g_2|^2}{\bar \nu_2} \right).
\end{equation}
The advantage of a double-pole over a single-pole $S$-matrix
parametrisation is that we can account for the interplay between a
potential resonance and a Feshbach resonance, which in principle can
radically change the scattering properties. This interplay is not only
important for the description of $^6$Li interactions, but also for
other atomic systems which have an almost resonant triplet potential,
such as bosonic $^{133}$Cs \cite{kokkelmans,leo} and $^{85}$Rb
\cite{kempen}.

In the many-body part of this paper, Section~\ref{hfbsection}, the
scattering properties are represented by a $T$-matrix instead of an
$S$-matrix. We have shown in the above that in our case there exists a
simple relation between the two, however, the definition for $T$ in
the many-body theory will be slightly different in order to give it
the conventional dimensions of energy per unit density:

\begin{equation}
  T(k)=\frac{2\pi \hbar ^{2}i}{mk}\left[ S(k)-1\right] .
\end{equation}

\subsection{Coupled square-well scattering} \label{doublewell}

\begin{figure}[h]
\begin{center}\
  \epsfxsize=80mm \epsfbox{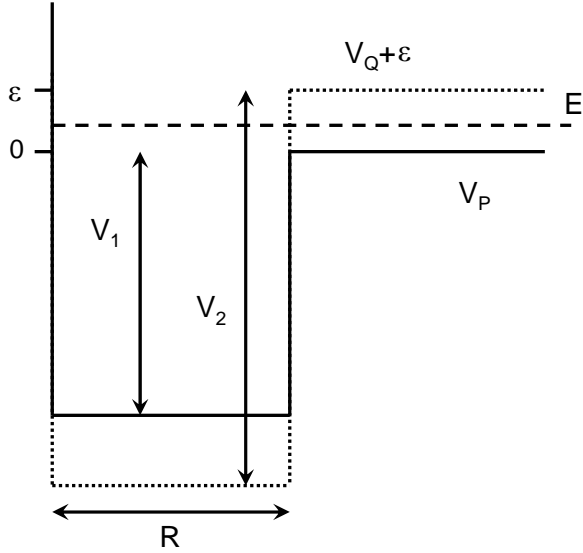}
\end{center}
\caption{
  Illustration of the coupled square well system. Outer region $r>R$:
  the solid line corresponds to the open channel potential $P$, and
  the dotted line to the closed channel potential $Q$. The wave
  functions are given by $u_P(r)\sim \sin k_P r$ and
  $u_Q(r)\sim\exp(-k_Qr)$, respectively. Inner region $r<R$: the solid
  and dotted lines correspond to the molecular potentials $V_1$ and
  $V_2$, respectively. The wave functions are given by $u_1(r)\sim
  \sin k_1 r$ and $u_2(r)\sim\sin(k_2r)$. The dashed line corresponds
  to the kinetic energy $E$ in the open channel. The
  wavevectors are defined as: $k_P=\sqrt{m E}/\hbar$,
  $k_Q=\sqrt{m (\epsilon-E)}/\hbar$, $k_1=\sqrt{m (E
      +V_1)}/\hbar$, $k_2=\sqrt{m (E
      +V_2-\epsilon)}/\hbar$. The detuning $\epsilon$ can be
  chosen such that a bound state of square-well potential $V_2$ enters
  the collision continuum, causing a Feshbach resonance in the open
  channel.} \label{fig3}
\end{figure}

In this subsection we describe the coupled-channels extension of a
textbook single-channel square-well scattering problem. One reason
that this model is interesting to study is because we can take the
limit of the potential range $R\rightarrow 0$, thus giving an explicit
representation of a set of coupled delta-function potentials which
simplifies the description in the many-body problem to follow.

The scattering equations for such a coupled system are written as

\begin{eqnarray}
  E\psi ^{P}({\bf r}) &=&\left[-\frac{\hbar^2}{m}\nabla^2_{\bf r}
  +V^{P}({\bf r})\right] \psi ^{P}({\bf
    r})+g({\bf r})\psi ^{Q}({\bf r}), \label{squarewellscateqns1} \\
  E\psi ^{Q}({\bf r}) &=&\left[-\frac{\hbar^2}{m}\nabla^2_{\bf r}+V^{Q}({\bf
      r})+\epsilon\right] \psi ^{Q}({\bf r} )+g^{\ast }({\bf r})\psi
  ^{P}({\bf r}),
\label{squarewellscateqns2}
\end{eqnarray}
with $\epsilon$ the energy-shift of the closed channel with respect to the
collision continuum and $E=\hbar^2k^2/m$ the relative kinetic energy
of the two colliding particles in the center of mass frame. The
coupled square well model encapsulates the general properties of two-body alkali
interactions. There we can divide the internuclear separation into two
regions: the inner region where the exchange interaction (the
difference between the singlet and triplet potentials) is much larger
than the hyperfine splitting, and the outer region where the hyperfine
interaction dominates. Here we make a similar distinction for the
coupled square wells. In analogy to the real singlet and triplet
potentials, we use for the inner region two artificial square-well
potentials labeled as $V_1$ and $V_2$. We take the coupling $g({\bf
  r})$ to be constant over the range of the square-well potentials
$r<R$, and to be zero outside this range (see Fig.~\ref{fig3}). Then
the problem can be simply solved by means of basis rotations at the
boundary $R$ giving rise to simple analytic expressions. For $r>R$, we
therefore consider one open channel and one closed channel, with
wavenumbers $k_{P}$ and $k_{Q}$. In analogy
with a real physical system, we can refer to the inner range channels
($r<R$) as a molecular basis, and the channel wave functions are just linear
combinations of the $u_1$ and $u_2$ wave functions. At the boundary
$R$, these wave functions have accumulated a phase $\phi_1=k_1R$ and
$\phi _2=k_2R$. The coupling strength is effectively given by the
basis-rotation angle $\theta$ for the scattering wave functions:
\begin{equation}
  \left(\begin{array}{c}u_P(R)\\u_Q(R)\end{array}\right)=
  \left(\begin{array}{cc}\cos \theta&-\sin \theta\\
   \sin \theta&\cos \theta\end{array}\right)
  \left(\begin{array}{c}u_1(R)\\u_2(R)\end{array}\right),
\end{equation}
allowing for an analytic solution of the scattering model. This leads
to the following expression for the $S$-matrix:

\begin{eqnarray}
  S&& =e^{-2ik_{P}R}\left[1- \right. \\
  &&(-2ik_{P}(k_{2}\cot \phi _{2}\cos ^{2}\theta+k_{Q}+k_{1}
  \cot \phi _{1}\sin ^{2}\theta ))/  \nonumber \\
  &&(k_{P}k_{Q}+k_{1}\cot \phi _{1}(k_{P}\sin
  ^{2}\theta-k_{Q}\cos ^{2}\theta)+  \nonumber \\
  &&\left. ik_{2}\cot \phi _{2}(k_{1}\cot \phi _{1}+ k_{P}\cos
    ^{2}\theta +k_{Q}\sin ^{2}\theta ))\right].\nonumber
\end{eqnarray}
An extention to treat more than two coupled potentials, which would be
required to model more than one resonance, is also straightforward.

The parameters of the two wells have to be chosen such that the
results of a real scattering calculation are reproduced for a given
physical system. In fact all the parameters are completely determined
from the field dependence of the scattering length, and all other
scattering properties, such as the energy-dependence of the scattering phase
shift, can then be derived. First we choose a range $R$, typically of the order
of an interatomic potential range (100 $a_0$) or less. Now we have only to
determine the set of parameters $V_{1}$, $V_{2}$ and $\theta $. The potential
depth $V_{1}$ is chosen such that the scattering length is equal to the
background scattering length $a_{\rm bg}$, while keeping $\theta=0$.
Also, $V_1$ should be large enough that the wavenumber $k_1$
depends weakly on the scattering energy. Then, we set $\theta$ to be
non-zero, and change the detuning until a bound state crosses
threshold, giving rise to a Feshbach resonance. The value of
$V_2$ is more or less arbitrary, but we typically choose it to be
larger than $V_1$.  Finally, we change the value of $\theta$ to give
the Feshbach resonance the desired width.

\begin{figure}[h]
\begin{center}\
  \epsfysize=60mm \epsfbox{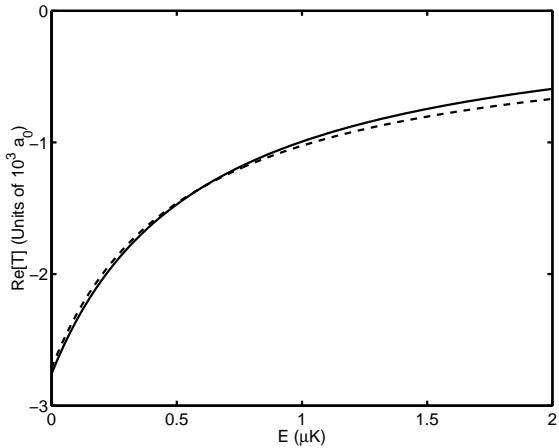}
\end{center}
\caption{Comparison of the real part of the $T$-matrix for coupled square-well
scattering (solid line) with a   potential range $R=1 a_0$, to Feshbach
scattering (dashed line), for   a detuning that yields a scattering length of
about -2750 $a_0$.   Similar good agreement is found for all detunings.}
\label{fig4} \end{figure}

We will later show that the resulting scattering properties converge
for $R\rightarrow 0$.  In Fig.~\ref{fig4} the coupled square-well
system is compared with the Feshbach scattering theory, for $^{40}$K
scattering parameters. Even despite the fact there is a strong
energy-dependence of the $T$-matrix, the two scattering
representations agree very well.

\subsection{Contact potential scattering and renormalization}
\label{cutoffscat}

In this subsection the Lippmann-Schwinger scattering equation is
solved for a resonance system with contact potentials. As in the
previous subsection, we make use of an open subspace that is coupled
to a closed subspace. The contact potentials are defined by:

\begin{eqnarray}
\label{VPcontact}
  V^{P}({\bf r}) &=&V^{P}\delta ({\bf r}), \\
  V^{Q}({\bf r}) &=&V^{Q}\delta ({\bf r}),  \nonumber \\
  g({\bf r}) &=&g \, \delta ({\bf r}),
\end{eqnarray}
with $\delta({\bf r})$ is the three-dimensional Dirac delta-function.
Here $V^{P}({\bf r})$ is the open channel potential with strength
$V^{P}$.  The function $V^{Q}({\bf r})$ is a closed channel potential
with strength $V^{Q}$, and $g({\bf r})$ is a coupling between the
closed and open channel with strength $g$.  The procedure of
renormalization relates the physical units ($a_{\rm bg}$, $\bar g_i$,
and $\bar \nu_i$) from Section~\ref{FBrep} to these parameters of the
contact potential scattering model for given momentum cutoff; a
relationship for which we will now obtain explicit expressions. The first
step is to solve again the scattering Eqns.~(\ref{squarewellscateqns1})
and (\ref{squarewellscateqns2}) for these contact potentials. As we
have seen in subsection \ref{FBrep}, we can formally solve the
bound-state equations, and make use of Eq.~(\ref{expansion}) to
expand the Green's function in bound-state solutions. In this case it
can be written as

\begin{equation}
  \psi ^{Q}({\bf r})=\sum_{i}\frac{\phi _{i}^{Q}({\bf r}) \int d^3r'
    \phi_{i}^{Q\ast}({\bf r'}) g^{\ast }({\bf r'}) \psi^{P}({\bf r'})}
  { E-\epsilon _{i}},
\end{equation}
with $\phi _{i}^{Q}({\bf r})$ a bound state solution and $\epsilon_i$
its eigenenergy. We now define an amplitude for the system to be in
this bound state that will later be useful in the mean-field
equations: $\phi _{i}=\langle \phi _{i}^{Q}|\psi ^{Q}\rangle $, and
together with the open channel equation and the definition $
g_{i}({\bf r})=g( {\bf r})\phi^Q_{i}({\bf r}) $, we get a new set of
scattering equations
\begin{eqnarray}
  \frac{\hbar ^{2}k^{2}}{m }\psi ^{P}({\bf r})
  &=&\left[-\frac{\hbar^2}{m}\nabla^2_{\bf r} +V^{P}({\bf r})\right] \psi
^{P}({\bf r})+\sum_i   g_{i}({\bf r})\phi_i,
\label{openscatnew} \\
\frac{\hbar^{2}k^{2}}{m }\phi _{i} &=&\nu_{i}\phi_{i}+\int d^{3}r' g_{i}^{\ast
  }({\bf r' })\psi ^{P}({\bf r'}). \label{amplitudeeq}
\end{eqnarray}
The energy-difference between the bound-state energy and the threshold of the
collision continuum is given by $\nu_{i}$. The open channel solution for
Eq.~(\ref{openscatnew}) can be formulated as
\begin{eqnarray}
  \psi ^{P}({\bf r}) &=&\chi({\bf r})-\frac{m }{4\pi \hbar
    ^{2}}\int d^{3}r^\prime \frac{e^{ik|{\bf r-r}^{\prime }|}}{|{\bf
r-r}^{\prime       }|}\left[ V^{P}({\bf r}
    ^{\prime })\psi ^{P}({\bf r}^{\prime })\right.   \nonumber \\
  &&+\sum_{i}\left. g_{i}({\bf r^\prime})\phi _{i}\right]  \\
  &=&\chi({\bf r})+f(\theta )\frac{e^{ikr}}{r},\quad \mbox{as $r\rightarrow
\infty$}.  \nonumber
\end{eqnarray}
Here $\chi({\bf r})$ is the unscattered wavefunction, and in the other
term we recognize the scattered part that is usually formulated in
terms of the scattering amplitude $f(\theta )$. The momentum
representation of this last line is~\cite{abrikosov}:
\begin{equation}
  \psi ^{P}({\bf p})=(2\pi )^{3}\delta ({\bf k-p})-\frac{4\pi f(k,p)}{
    k^{2}-p^{2}+i\delta }.  \label{abrikosovscatamp}
\end{equation}
Combining Eq.~(\ref{abrikosovscatamp}) with our expression for the
scattering amplitude we find

\begin{eqnarray}
  -\frac{4\pi \hbar ^{2}}{m }f(k,k^{\prime })
  &=&V^{P}+\frac{1}{(2\pi )^{3}} V^{P}\int d^{3}p\frac{-\frac{4\pi \hbar
      ^{2}}{m }f(k,p)}{\frac{\hbar ^{2}k^{2}}{
      m }-\frac{\hbar ^{2}p^{2}}{m }+i\delta }  \nonumber \\
  &&+\sum_{i}g_{i}\phi _{i}.  \label{scatlengthkspace}
\end{eqnarray}
The typical temperature range of a system we are interested in will only
allow for elastic $s$-wave scattering, therefore the scattering amplitude has no
angular dependence, and incoming and outgoing wavenumbers are the same, i.e.
$k=k'$. The scattering amplitude can then be simply linked to the
$T$-matrix via the relation $T(k)=-(4\pi \hbar ^{2}/m )f(k)$. The integral has
a principal-value part, and the integration ranges from zero to a momentum
cutoff $K $. Eq. (\ref{scatlengthkspace}) then has as solution

\begin{eqnarray}
  T(k) &=&V^{P}-\frac{V^{P}m}{2\pi^{2}\hbar
    ^{2}}T(k)\left[ K- {\rm arctanh}
    \frac{k}{K}+\frac{i\pi }{2}k\right]   \nonumber \\
  &&+\sum_{i}g_{i}\phi _{i}.  \label{lippman1}
\end{eqnarray}
This is a variant of the Lippmann-Schwinger equation. The
closed channel scattering solutions are now used to eliminate the
amplitude functions $\phi_{i}$. In Fourier space,
Eq.~(\ref{amplitudeeq}) has the form
\begin{equation}
  \frac{\hbar ^{2}k^{2}}{m }\phi _{i}=\nu _{i}\phi_{i}+g_{i}^{\ast
    }\frac{1}{(2\pi )^{3}}\int \psi ^{P}({\bf p})d^{3}p.
\end{equation}
After substitution of Eq.~(\ref{abrikosovscatamp}) the expression for
$\phi _{i}$ is linked to the $T$-matrix:
\begin{equation}
  \phi _{i}=\frac{g_{i}^{\ast }\left( 1-\frac{m}{2\pi
        ^{2}\hbar ^{2}}T(k)\left[ K-{\rm arctanh}
        \frac{k}{K}+\frac{i\pi }{2}k\right] \right) }{\frac{\hbar
      ^{2}k^{2}}{m } -\nu _{i}}.
\end{equation}
Eliminating $\phi_i$ from Eq.~(\ref{lippman1}) gives a complete expression for
the Lippmann-Schwinger equation

\begin{eqnarray}
  &&T(k) =V^{P}-\frac{V^{P}m}{2\pi^{2}\hbar^{2}}
  T(k)\left[ K-
    {\rm arctanh}\frac{k}{K}+\frac{i\pi }{2}k\right]   \nonumber \\
  &&+\sum_{i} \frac{|g_{i}|^{2}\left( 1-\frac{1}{2\pi ^{2}}\frac{m
        }{\hbar ^{2}}T(k)\left[ K-{\rm arctanh}\frac{k}{K}+\frac{i\pi
          }{2}k\right] \right) }{\frac{\hbar ^{2}k^{2}}{m }-\nu
    _{i}}.  \label{lippman2}
\end{eqnarray}

Similar to the Feshbach and coupled square-well problems, the
$k\rightarrow 0$ behavior of $T(k)$ should reproduce the scattering
length, and, the result should not depend on the arbitrary momentum
cutoff $K$. For an analytic expression of the scattering length, we
conveniently use the Feshbach representation. A comparison between the
latter and the expression for the scattering length $a$ that results from
solving Eq.~(\ref{lippman2}), tells us how to relate the coupling constants
for contact-scattering to the Feshbach coupling constants. By making
use of the definitions $\Gamma =(1-\alpha U)^{-1}$,
$\alpha=mK/(2\pi^2\hbar^2)$, and $U=4\pi \hbar^2 a_{\rm bg}/m$, we
find the very concise relations

\begin{equation}
V^{P} =\Gamma U,
\label{ren1}
\end{equation}
which is valid also in the case where no resonance is present, and in
addition

\begin{eqnarray}
  g_{1} &=&\Gamma \bar g_{1}, \\
  \nu _{1} &=&\bar \nu_1 + \alpha g_1 \bar g_1.
\end{eqnarray}
for the open channel potential and the first resonance. For the second
resonance, if present, we find
\begin{eqnarray}
  g_{2} &=& \frac{\bar g_2}{\alpha \bar g_1^2/\bar \nu_1+\Gamma^{-1}}, \\
  \nu _{2} &=&\bar \nu_2 + \alpha g_2 \bar g_2. \label{ren5}
\end{eqnarray}
Obviously, our approach an be systematically extended further, order by order,
to give an arbitrarily accurate representation of the microscopic scattering
physics.

These expressions we refer to as the renormalizing equations of the
resonance theory since they remove the ultra-violet divergence which
would otherwise appear in the field-equations. Any many-body theory
based on contact scattering around a Feshbach resonance will need to
apply these expressions in order to renormalize the theory. These equations
(\ref{ren1})-(\ref{ren5}) therefore represent one of the major results of this
paper.

\begin{figure}[h]
\begin{center}
  \epsfysize=60mm \epsfbox{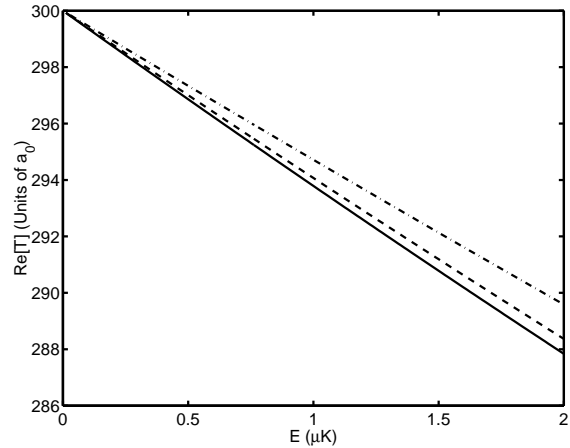}
\end{center}
\caption{
Comparison of the real part of the $T$-matrix for coupled square-well scattering
for three different values of the potential range: $R=100 a_0$ (dash-dotted
line), $R=30 a_0$ (dashed line), $R=1 a_0$ (solid line). The interaction
parameters for $^{40}$K have been used here, and the magnetic field is chosen
such that a scattering length of $a=300a_0$ is obtained. Also plotted is the
$T$-matrix for contact scattering, which clearly agrees very well as it
coincides with the solid line of the double-well scattering.}
\label{fig5}
\end{figure}

In Fig.~\ref{fig5} the $T$-matrix as a function of energy is shown
for contact scattering, in comparison with the square-well scattering for
different values of the potential range. The contact-scattering model is
demonstrated to be the limiting case of the coupled square well system when
$R\rightarrow~0$.

\subsection{Discussion of different models}

\begin{figure}[h]
\begin{center}
 \epsfxsize=80mm \epsfbox{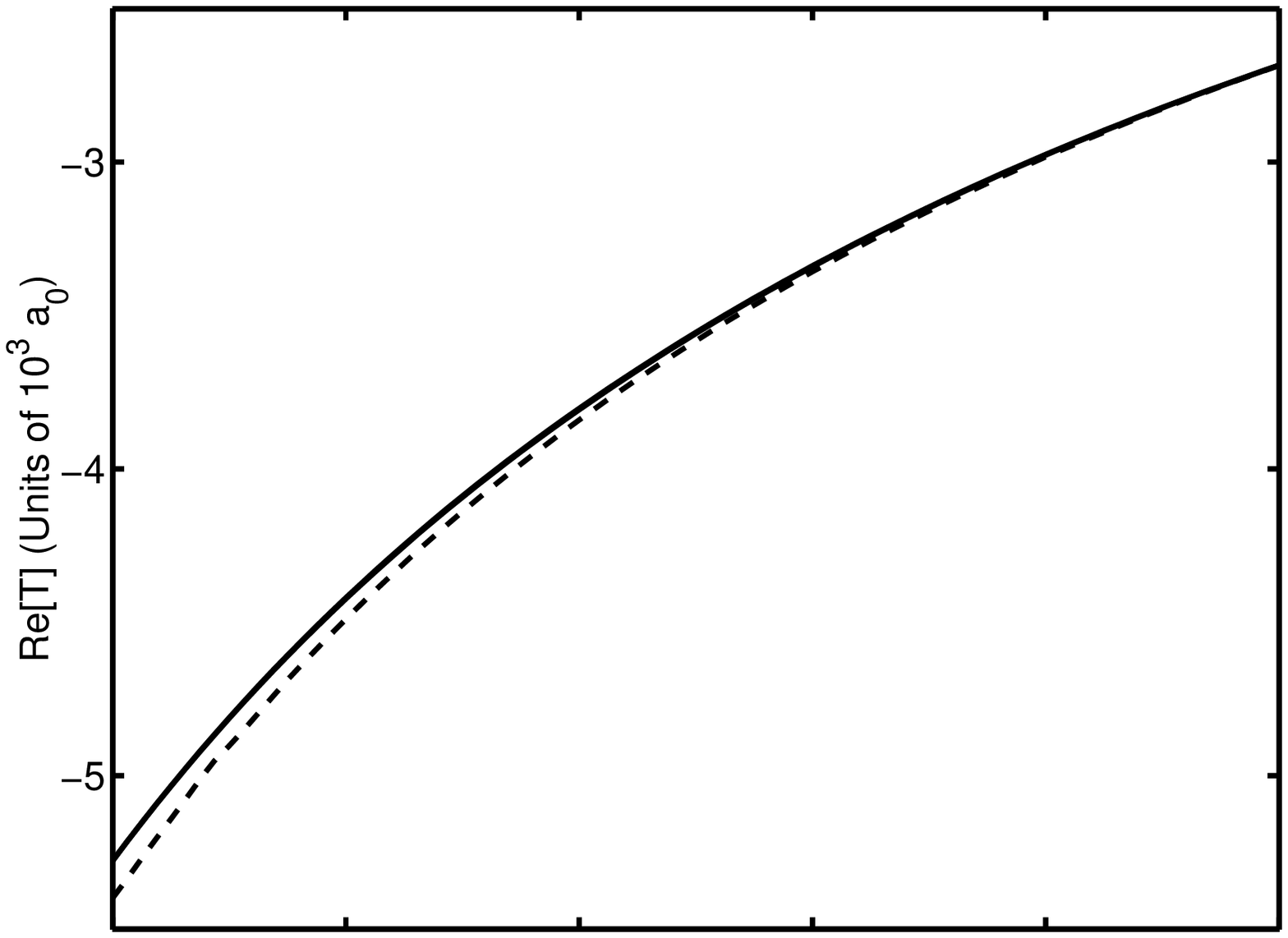}
 \epsfxsize=80mm \epsfbox{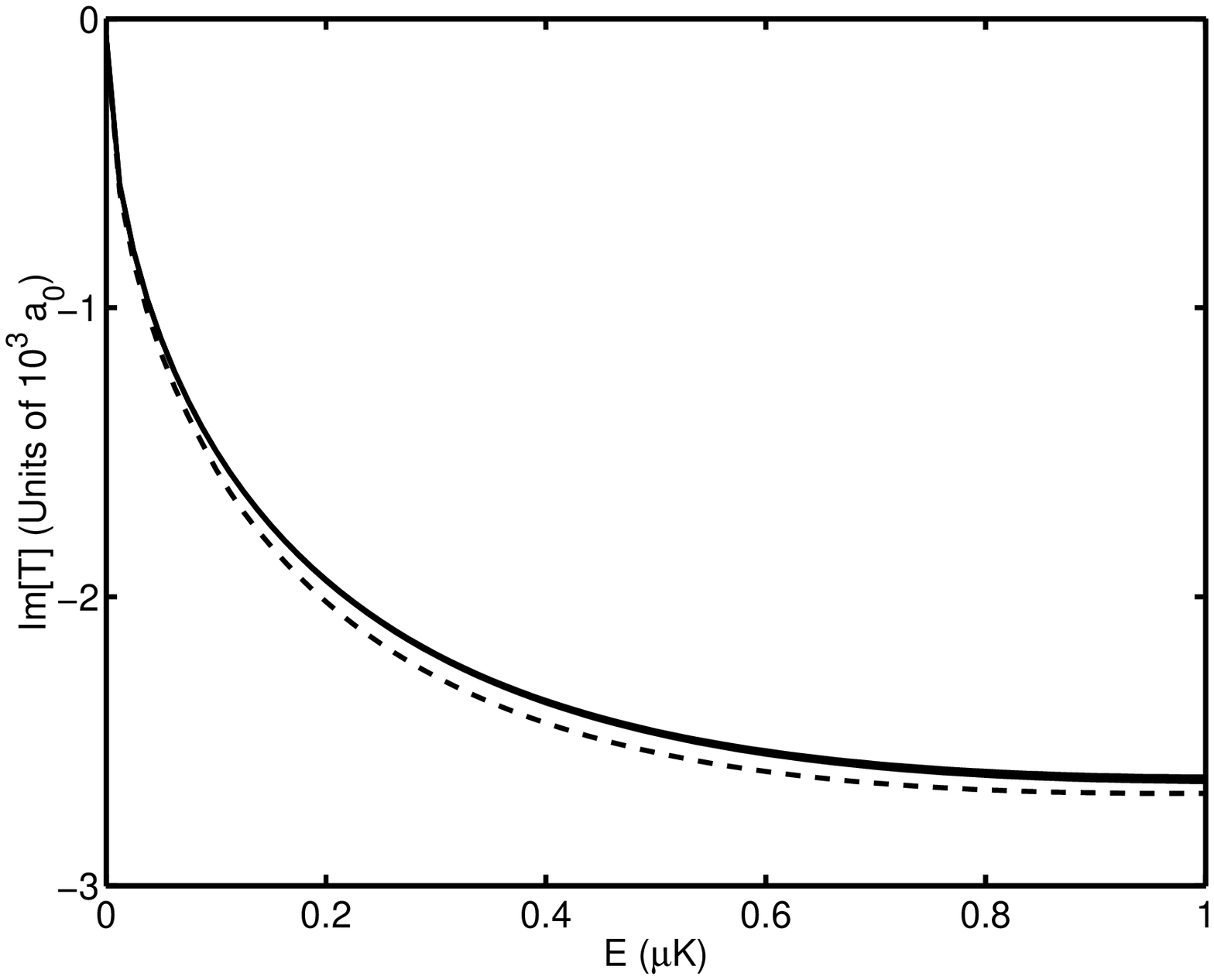}
\end{center}
\caption{(a) Real part of the $T$-matrix as a function of collision
  energy, for the Feshbach model and the cutoff model (overlapping
  solid lines), and for a coupled channels calculation (dashed line).
  The atomic species considered is $^6$Li, for atoms colliding in the
  $(1/2,-1/2) +(1/2,1/2)$ channel. (b) Same as (a) for the imaginary
  part.}
\label{fig6}
\end{figure}

In subsection \ref{cutoffscat} it has been shown that the resonance contact
scattering representation is the limiting case of the coupled
square-well system, when the range of the potentials is taken to zero.
Also, in subsection \ref{doublewell} it has been shown that the
double-well system is in good agreement with the Feshbach scattering
theory. Now we will show how well these scattering representations
agree with the full numerical coupled channels calculation~\cite{servaaspriv}.
In Fig.~\ref{fig6} we show the real and imaginary parts of the $T$-matrix
applied to the case of $^{6}$Li, and compare the cutoff and Feshbach scattering
representations to a full coupled channels calculation. The agreement
is surprisingly good, and holds basically for all magnetic fields (i.e. similar
agreement is found at all detunings).

In this section we have discovered a remarkable fact that even a complex
system including internal structure and resonances can be simply described with
contact potentials and a few coupling parameters. This was known for
off resonance scattering where only a single parameter (the scattering length)
is required to encapsulate the collision physics at very low temperature.
However, to our knowledge this has not been pointed out before for the resonance
system, where an analogous parameter set is required to describe a system where
the scattering length may even pass through infinity. We have shown in a very
concise set of formulas how to derive the resonance parameters associated with
contact potentials. This result is important for the incorporation of the 
two-body scattering in a many-body system, as we will show later in this paper.

Other papers have also proposed a simple scattering model to reproduce coupled
channels calculations \cite{houbiers,vogels}. In these papers real potentials
are used, and they give a fair agreement. Here, however, we use models that need 
input from a coupled channels calculation to give information about the
positions of the bound states and the coupling to the closed channels. All this
information can be extracted from a plot of the scattering length as a function
of magnetic field.

\section{Many-body resonance scattering} \label{hfbsection}

We will now proceed to a many-body description of resonance
superfluidity and connect it to our theory of the two-body scattering
problem described earlier. This section explains in detail the similar
approach in our papers devoted to resonance superfluidity in potassium
\cite{holland,chiofalo}. The general methods of non-equilibrium
dynamics has been described in \cite{peletminskii} and we have applied
them in the context of condensed bosonic fields
\cite{walser599,walser999}.

In the language of second quantization, we describe the many-body
system with fermionic fields $\annhilate{\psi}{\sigma}(\boldvec x)$
which remove a single fermionic particle from position ${\boldvec x}$
in internal electronic state $\sigma$, and molecular bosonic fields
$\annhilate{\phi}{i}(\boldvec x)$ which annihilate a composite bound
two-particle excitation from space-point ${\boldvec x}$ in internal
configuration $i$. These field operators and their adjoints satisfy
the usual fermionic anti-commutation rules
\begin{eqnarray}
  \left \{ \annhilate{\psi}{\sigma_1}({\boldvec x}_1),
    \create{\psi}{\sigma_2}({\boldvec x}_2) \right \}&=&
\delta({\boldvec x}_1-{\boldvec x}_2)\,
\delta _{\sigma_1 \sigma_2}\equiv\delta_{12}, \nonumber\\
  \left \{\annhilate{\psi}{\sigma_1}({\boldvec x}_1),
  \annhilate{\psi}{\sigma_2}({\boldvec x}_2) \right \} &=&0,
\end{eqnarray}
and bosonic commutation rules
\begin{eqnarray}
\left [ \hat{\phi}_{i_1}^{\phantom\dag}({\boldvec x}_1),
    \hat{\phi}_{i_2}^{\dag}({\boldvec x}_2) \right ]&=&
  \delta({\boldvec x}_1-{\boldvec x}_2)\,
\delta_{i_1 i_2}\equiv \delta_{12},\nonumber \\
\left[\hat{\phi}_{i_1}^{\phantom\dag}({\boldvec x}_1),
\hat{\phi}_{i_2}^{\phantom\dag}({\boldvec x}_2)\right]
 &=& 0,
\end{eqnarray}
respectively.  In here and the following discussion, we will also try
to simplify the notational complexity by adopting the notation
convention of many-particle physics. This means, we will identify the
complete set of quantum numbers uniquely by its subscript index, i.e.,
$\{{\boldvec x}_1,\sigma_1\} \equiv 1$. If only the position coordinate
is involved, we will use bold face ${\boldvec x}_2\equiv \bz $.

In the double resonance case of lithium, we have to distinguish
only two internal atomic configurations for the free fermionic single
particle states $\sigma=\{\uparrow, \downarrow\}$ and we need at most two
indices $i=\{1,2\}$ to differentiate between the bosonic molecular
resonances.

The dynamics of the multi-component gas is governed by a total system
Hamiltonian $\hat{H}=\hat{H}_0+\hat{H}_1$, which consists of the free
evolution Hamiltonian $\hat{H}_0$ and the interactions $\hat{H}_1$
between atoms and molecules. We assume that the free dynamics of the
atoms and molecules is determined by their kinetic and potential
energies in the presence of external traps, which is measured relative
to the energy $\mu$ of a co-rotating reference system. Thus, we define
\begin{eqnarray}
\label{freefermi}
H_{\sigma}(\boldvec x) &=&-\frac{\hbar ^{2}}{2m}\nabla
^{2}+V_{\sigma}(\boldvec x)-\mu, \\
H_{i}^{m}(\boldvec x) &=&-\frac{\hbar ^{2}}{2M}\nabla ^{2}+V_{i}^{m}(
\boldvec x)+\nu _{i}-\mu_m.
\end{eqnarray}
Here, $m$ denotes the atomic mass as used previously, $M=2 \, m$ is the
molecular mass, $\mu_m=2 \, \mu$ is the energy offset of the molecules with
respect to the reference system, $V_{\sigma}(\boldvec x)$ are external
spin-dependent atomic trapping potentials, and $V_{i}^{m}(\boldvec x)$ are the
external molecular trapping potentials. The molecular single-particle
energy has an additional energy term $\nu _{i}$ that accounts for the
detuning of the molecular state $i$ relative to the threshold of the collision
continuum.

The binary interaction potential $V^{P}({\boldvec x}_1-{\boldvec
  x}_2)$ accounts for the non-resonant interaction of spin-up and spin-down
fermions, and coupling potentials $g_{i}({\boldvec x}_1-{\boldvec x}_2)$ convert
free fermionic particles into bound bosonic molecular excitations. Thus, we find
for the total system Hamiltonian of the atomic and molecular fields:
\begin{eqnarray}   \hat{H}&=&\hat{H}_0+\hat{H}_1,
\label{fullpot}
\end{eqnarray}
where the free $\hat{H}_0$ and interaction contributions $\hat{H}_1$
are defined as
\begin{eqnarray}
\label{H0op}
\hat{H}_0&=&\int d{\be}
   \sum_{\sigma} \create{\psi}{\sigma}(\be)H_{\sigma}(\be)
  \annhilate{\psi}{\sigma}({\be})\nonumber\\
 &+&\int d{\be} \sum_{i} \create{\phi}{i}(\be)H_{i}^{m}(\be)
  \annhilate{\phi}{i}(\be),\\
\label{H1op}
\hat{H}_1&=&
  \int d{\be} d{\bz}\left\{ \create{\psi}{\uparrow}(\be)
    \create{\psi}{\downarrow }(\bz)
    V^{P}(\be-\bz)
    \annhilate{\psi}{\downarrow}(\bz)
    \annhilate{\psi}{\uparrow }(\be)\right. \nonumber\\
  \lefteqn{+\sum_{i} \left. \left [
        \create{\phi}{i}(\frac{{\be}+{\bz}}{2})
        g_{i}^{\ast}({\be}-{\bz}) \annhilate{\psi}{\downarrow}({\bz})
        \annhilate{\psi}{\uparrow}({\be})+{\rm H.c.}\right]\right\}.}&&
  \nonumber\\
\end{eqnarray}
Here, H.c.~denotes the Hermitian conjugate.  In the present picture,
we deliberately neglect the interactions among the molecules.  Several
other papers have treated a Feshbach resonance in a related
way~\cite{timmermans,abeelen2,holland2,kokkelmans2}.

In order to derive dynamical Hartree-Fock-Bogoliubov (HFB) equations
from this Hamiltonian, we also need to define a generalized density
matrix to describe the state of the fermionic system \cite{blaizot}
and an expectation value for the bosonic molecular field.
The elements of the $4\times4$ density matrix ${\cal G}$ are given by
\begin{eqnarray}
  {\cal G}_{pq}(\be \bz)
    &=&\langle
    \create{A}{q}({\boldvec x}_2)
    \annhilate{A}{p}({\boldvec x}_1)
    \rangle,\\
  \hat{A}({\boldvec x})&=&\left(
\annhilate{\psi}{\uparrow}({\boldvec x}),
\annhilate{\psi}{\downarrow}({\boldvec x}),
\create{\psi}{\uparrow }({\boldvec x}),
\create{\psi}{\downarrow }({\boldvec x})
\right)^\top,
\end{eqnarray}
and symmetry broken molecular fields are defined as
\begin{eqnarray}
\label{phidef}
\phi_i(\be)&=&\langle \annhilate{\phi}{i}({\boldvec x}_1)\rangle.
\end{eqnarray}
As usual, we define the quantum averages of an arbitrary operator ${\cal \hat
O}$ with respect to a many-body density matrix $ {\boldvec \rho}$ by $\langle
{\cal \hat O} \rangle={\rm   Tr}[{\cal \hat O} {\boldvec \rho}]$, and we
calculate higher order correlation functions by a Gaussian factorization
approximation known as Wick's theorem \cite{blaizot}. The structure of the
$4\times4$ density matrix
\begin{eqnarray}
\label{Gdef}
{\cal G}(\be\bz)&=&
\left(
  \begin{array}{cc}
   {\cal G_N}(\be\bz) &{\cal G_A}(\be\bz)\\
-{\cal G_A}(\be\bz)^\ast& \delta_{\be\bz}-{\cal G_N}(\be\bz)^\ast
  \end{array}
\right),
\end{eqnarray}
is very simple, if one recognizes that it is formed out of a
$2\times2$ single particle density matrix ${\cal G_N}$, a pair
correlation matrix ${\cal G_A}$ and obviously the vacuum fluctuations
$\delta_{\be\bz}$. The single particle submatrix is given by
\begin{equation}
 {\cal G_N}(\be\bz)=
\left(
  \begin{array}{cc}
   {\cal G}_{n\uparrow}(\be\bz) &{\cal G}_{m}(\be\bz)\\
   {\cal G}_{m}(\bz\be)^\ast& {\cal G}_{n\downarrow}(\be\bz)
  \end{array}
\right),
\end{equation}
where ${\cal G}_{n\sigma}(\be\bz)= \langle \create{\psi}{\sigma
  }({\boldvec x}_2) \annhilate{\psi}{\sigma}( {\boldvec x}_1)\rangle $
is the density of spin-up and down particles and ${\cal
  G}_{m}(\be\bz)=\langle \create{\psi}{\downarrow }({\boldvec x}_2)
\annhilate{\psi}{\uparrow}( {\boldvec x}_1)\rangle $ denotes a
cross-level coherence, or ``magnetization'' between the states.  The
pair-correlation submatrix ${\cal G_A}$ is defined analogously as
\begin{equation}
{\cal G_A}(\be\bz)=
\left(
  \begin{array}{cc}
    {\cal G}_{a\uparrow}(\be\bz) &  {\cal G}_{p}(\be\bz) \\
    -{\cal G}_{p}(\bz\be) &  {\cal G}_{a\downarrow}(\be\bz)
 \end{array}
\right),
\end{equation}
where ${\cal G}_{a\sigma}(\be\bz)=\langle
\annhilate{\psi}{\sigma}({\boldvec x}_2)
\annhilate{\psi}{\sigma}({\boldvec x}_1)\rangle $ is an anomalous
pairing field within the same level and the usual cross-level pairing
field of BCS theory is defined in here as ${\cal G}_{p}(\be\bz)=\langle
\annhilate{\psi}{\downarrow }({\boldvec x}_2)
\annhilate{\psi}{\uparrow}({\boldvec x}_1)\rangle$.

\subsection{General dynamic Hartree-Fock-Bogoliubov equations of motion}
From these physical assumptions about the system's Hamiltonian
Eq.~({\ref{fullpot}) and the postulated mean fields $\phi_i$ of
  Eq.~({\ref{phidef}) and ${\cal G}$ of Eq.~({\ref{Gdef}), one can now
      derive kinetic equations for the expectation values $\langle
      {\cal O} \rangle$ for an operator ${\cal O}$ by a
      systematic application of Heisenberg's equation
\begin{equation}
  i\hbar \frac{d}{dt} \hat{{\cal O}}=[\hat{{\cal O}},\hat{H}],  \label{heis}
\end{equation}
and Wick's theorem.

The first order kinetic equation for the Hermitian density matrix
${\cal G}$ has the general form of a commutator and the time-evolution
is determined by a Hermitian self-energy matrix
$\Sigma=\Sigma^0+\Sigma^1$.  In general, one finds
\begin{eqnarray}
\label{HFB}
i \hbar \frac{d}{dt}{\cal G}(\be\bd)&=&
\int d\bz\left[\Sigma(\be\bz)\,
{\cal G}(\bz\bd)-{\cal G}(\be\bz)\Sigma(\bz\bd)\right] ,\\
i \hbar  \frac{d}{dt}\phi_i(\bd)&=&H_{i}^{m}(\bd)\, \phi_i(\bd)\\
\lefteqn{+\int d\be d\bz\, \delta(\frac{\be+\bz}{2}-\bd)g_i^{\ast}(\be-\bz)
{\cal G}_{p}(\be\bz).} &&\nonumber
\end{eqnarray}
First, the free evolution $\Sigma^0$ is obviously related to
the single particle Hamiltonians of Eq.~(\ref{H0op}). In complete
analogy to the generalized density matrix, it has a simple $4\times4$
structure
\begin{eqnarray}
\Sigma^0(\be\bz)&=& \left(
    \begin{array}{cc}
      \Sigma_{{\cal N}}^0(\be\bz) & 0 \\
      0&  -{\Sigma_{{\cal N}}^0}(\be\bz)^\ast
    \end{array}
  \right),
\end{eqnarray}
which can  be factorized into $2\times2$ submatrices as
\begin{eqnarray}
  \Sigma_{{\cal N}}^0(\be\bz)&=&\delta_{\be\bz}
  \left(
    \begin{array}{cc}
      H_{\uparrow}(\be) & 0\\
      0& H_{\downarrow}(\be)
    \end{array}
  \right).
\end{eqnarray}
Secondly, one obtains from the interaction Hamiltonian of
Eq.~(\ref{H1op}) the first order self-energy $\Sigma^1$ as
\begin{eqnarray}
\Sigma^1(\be\bz)&=&
\left(
  \begin{array}{cc}
   \Sigma_{{\cal N}}^1(\be\bz) &  \Sigma_{{\cal A}}^1(\be\bz) \\
   - \Sigma_{{\cal A}}^1(\be\bz)^\ast &  -\Sigma_{{\cal N}}^1(\be\bz)^\ast
  \end{array}
\right).
\end{eqnarray}
The normal potential matrix $\Sigma_{{\cal N}}^1$ has the usual
structure of direct contributions [i.e. local Hartree potentials
proportional to $\delta_{\be\bz}$] and exchange terms [i.e.  non-local
Fock potentials proportional to $V^P(\be-\bz)$]:
\begin{eqnarray}
  \Sigma_{{\cal N}}^1(\be\bz)&=&
\int d\bv V^P(\bz-\bv)
\left(
  \begin{array}{cc}
   \delta_{\be\bz} {\cal G}_{n\downarrow}(\bv\bv)&
   -\delta_{\be\bv} {\cal G}_{m}(\be\bz)\\
   -\delta_{\be\bv}{\cal G}_{m}(\bz\be)^\ast&
   \delta_{\be\bz}{\cal G}_{n\uparrow}(\bv\bv)
  \end{array}
\right).\nonumber\\
\end{eqnarray}
The zeros that appear in the diagonal of
the anomalous coupling matrix
\begin{eqnarray}
 \Sigma_{{\cal A}}^1(\be\bz)&=&
\left(
  \begin{array}{cc}
    0 &\Delta(\be\bz)\\
   -\Delta(\bz\be)&0
\end{array}
\right),
\end{eqnarray}
reflect the fact that there is no low energy ($s$-wave) interaction
between same spin particles due to the Pauli exclusion principle. The
off-diagonal element defines a gap function as
\begin{eqnarray}
\Delta(\be\bz)&=& V^P(\be-\bz)\,{\cal G}_{p}(\be\bz)
+\sum_i  g_i(\be-\bz)\,\phi_i(\frac{\be+\bz}{2}).\nonumber\\
\end{eqnarray}
\subsection{The homogeneous limit and the contact potential
 approximation}

In this section, we will apply the general HFB equations of motion
[Eq.~(\ref{HFB})] to the case of a spatially homogeneous isotropic
system.  Furthermore, we will approximate the finite range interaction
potentials $V^P({\boldvec x}_1-{\boldvec x}_2)$ and $g_i({\boldvec
  x}_1-{\boldvec x}_2)$ by the contact approximation as introduced in
Eq.~(\ref{VPcontact}), and assume equal populations for spin-up and spin-down
atoms.

Spatial homogeneity implies that a physical system is translationally
invariant. Thus, any single particle field must be constant in space
and any two-particle quantity or pair-correlation function can depend
on the coordinate difference only:
\begin{eqnarray}
  \phi_i({\boldvec x})&=&\phi_i({\boldvec 0})\equiv\phi_i,\\
  {\cal G}({\boldvec x}_1, {\boldvec x}_2)&=&
  {\cal G}({\boldvec x}_1-{\boldvec x}_2)={\cal G}({\boldvec r}).
\end{eqnarray}
This assumption implies also that there can be no external trapping
potentials present, i.e. $V_\sigma({\boldvec x})=V_i^m({\boldvec
  x})=0$, as this would break the translational symmetry.

Furthermore, we want to consider a special situation when there is no population
difference in spin-up and spin-down particles ${\cal G}_{n}(r=|{\boldvec
r}|)={\cal G}_{n\sigma}({\boldvec r})$, there excists no cross-level
coherence or ``magnetization'' ${\cal G}_m(r)=0$, and the anomalous
pairing field ${\cal G}_a(r)=0$. It is important to note that this
special scenario is consistent with the full evolution equation and, on the
other hand, leads to a greatly simplified sparse density matrix:

\begin{eqnarray}  {\cal G}(\be\bz)&=& \left(
\begin{array}{cccc}
  {\cal G}_{n}(r) & 0 & 0 &  {\cal G}_{p}(r)\\
  0 & {\cal G}_{n}(r) & - {\cal G}_{p}(r) & 0\\
  0 & - {\cal G}_{p}^\ast(r) & \delta({\boldvec r})- {\cal G}_{n}(r) & 0\\
  {\cal G}_{p}^\ast(r)  & 0 & 0 & \delta({\boldvec r})- {\cal G}_{n}(r)
\end{array}
\right),\nonumber\\
\end{eqnarray}
where $r=|{\boldvec r}|=|\be-\bz|$.
Similarly, one finds a
translationally invariant self-energy $\Sigma(\be\bz)=\Sigma(\be-\bz)$
with
\begin{eqnarray}
\Sigma(\be\bz)&=&\delta_{\be\bz}\left(
\begin{array}{cccc}
  \Sigma(\be) & 0 & 0 & \Delta  \\
  0 & \Sigma(\be) & -\Delta  & 0 \\
  0 & -\Delta^\ast & -\Sigma(\be) & 0 \\
  \Delta ^{\ast } & 0 & 0 & -\Sigma(\be)
 \end{array}
\right),
\end{eqnarray}
and  $\Sigma({\boldvec x})=-\hbar ^{2}/(2m )\nabla_{{\boldvec x}}
^{2}-\mu+V^{P}{\cal G}_{n}(0)$ and a complex energy gap $\Delta
=V^{P}{\cal G}_{p}(0)+\sum_{i} g_{i} \,\phi_{i}$.  These assumptions
lead to a significant simplification of the HFB equations.

The structure of the HFB equations can be elucidated further by
separating out the bare two-particle interactions from the many-body
contributions.  One can achieve this by splitting the self-energy into
the kinetic energy and mean-field shifts
$\Sigma=\Sigma^{0}+\Sigma^{1}$, and by separating the density matrix
into the vacuum contribution ${\cal G}^0$ [proportional to
$\delta(\boldvec r)$] and the remaining mean-fields ${\cal G}={\cal
  G}^{0}+{\cal G}^{1}$:
\begin{eqnarray}
\label{splitequation1}
  i\hbar \frac{d}{dt}{\cal G}^{1}&-&[\Sigma^{0},{\cal G}^{1}]
  -[\Sigma^{1}, {\cal G}^{0}]=[\Sigma^{1},{\cal G}^{1}],\\
\label{splitequation2}
  i \hbar  \frac{d}{dt}\phi_i&=&(\nu_i-\mu_m)\,\phi_i+g_i^\ast\,{\cal G}_{p}(0).
\end{eqnarray}
 In this fashion, we can now identify the
physics of resonance scattering of two particles in vacuo [left hand
side of Eq.~(\ref{splitequation1})] from the many-body corrections due to
the presence of a medium [right hand side of Eq.~(\ref{splitequation1})]

In the limit of very low densities, we can ignore many-body effects and
rediscover  Eqs.~(\ref{openscatnew}) and (\ref{amplitudeeq}) of subsection
\ref{cutoffscat}, but given here in a time-dependent form.  They
describe the scattering problem that we have solved already:
\begin{eqnarray}
  i\hbar \frac{d}{dt}{\cal G}_{n}({\boldvec r}) &=&0,\\
\label{Gp1}
  i\hbar \frac{d}{dt}{\cal G}_{p}({\boldvec r}) &=&
\left[ -\frac{\hbar^{2}}{m}\nabla^{2}_{\boldvec r}-2\mu
+V^{P}({\boldvec r})\right]
{\cal G}_{p}({\boldvec r})\nonumber\\
&&+\sum_{i}g_{i}({\boldvec r})\,\phi_{i}, \\
i\hbar \frac{d}{dt}\phi_{i}&=&(\nu_i-\mu_m)\, \phi_i
+ g_{i}^\ast\,{\cal G}_{p}(0). \label{phi1}
\end{eqnarray}

The scattering solution of Eqs.~(\ref{Gp1}) and (\ref{phi1}) is `summarized'
by the energy-dependent two-body $T$-matrix, which we have discussed in the
preceeding sections. In order to incorporate the full energy-dependence of the
scattering physics, we propose to upgrade the direct
energy-shift $V^{P}{\cal G}_{n}(\boldvec r)$ to $\langle T^{\rm   Re}(k)\rangle
{\cal G}_{n}(\boldvec r)$, where $\langle T^{\rm   Re}(k)\rangle$ represents the
real part of the two-body $T$-matrix, and $\langle\ldots \rangle$ denotes
two-particle thermal averaging over a Fermi-distribution. A detailed calculation
of the proper upgrade procedure will be presented in a forthcoming publication.

The translationally invariant HFB Eqns.~(\ref{splitequation1}) and
(\ref{splitequation2}) are best analyzed in momentum-space. Thus, we will
introduce the Fourier transformed field-operators $\hat{a}_{{\smallvec
k}\sigma}$ by

\begin{eqnarray}
  \annhilate{\psi}{\sigma}(\boldvec x)&=&\sum_{\smallvec k}
\frac{e^{-i \smallvec k \smallvec x}}{\sqrt{\Omega}}
\annhilate{a}{\smallvec{k}\sigma},
\end{eqnarray}
where $\Omega$ is the quantization volume. If we define the Fourier components
of the translationally invariant mean fields as

\begin{eqnarray}
{\cal G}({\boldvec r})&=&
{\cal G}({\boldvec x}_1-{\boldvec x}_2)=\sum_{\smallvec k}
e^{-i \smallvec k( \smallvec x_1-\smallvec x_2)}\,
{\cal G}({\boldvec k}),
\end{eqnarray}
we obtain the following relations between the real-space density of
particles $n$ (the same for both spins) and the real-space density of
particle pairs $p$
\begin{eqnarray}
  n&=&{\cal G}_{n\sigma}({\boldvec r}=0)= \sum_{\smallvec k}{\cal
    G}_{n\sigma}({\boldvec k})= \frac{1}{\Omega} \sum_{\smallvec k}
  \langle\create{a}{\smallvec k\sigma}
  \annhilate{a}{\smallvec k\sigma}\rangle,\\
  p&=&{\cal G}_{p}({\boldvec r}=0)=\sum_{\smallvec k} {\cal
    G}_{p}({\boldvec k})=\frac{1}{\Omega}\sum_{\smallvec
    k}\langle\annhilate{a}{-\smallvec
    k\downarrow}\annhilate{a}{\smallvec k\uparrow}\rangle.
\end{eqnarray}
The Fourier-transformed HFB equations are now local in momentum-space
\begin{eqnarray}
i \hbar \frac{d}{dt}{\cal G}\boldvec k)&=&
\left[\Sigma(\boldvec k),{\cal G}(\boldvec k)\right],\\
i\hbar \frac{d}{dt}\phi_{i}&=&(\nu_i-\mu_m)\, \phi_i
+ g_{i}^\ast\,p,
\end{eqnarray}
and the self-energy is given by
\begin{equation}
  \Sigma({\boldvec k})=\left(\begin{array}{cccc}
      \Sigma_k & 0 & 0 & \Delta\\
      0 & \Sigma_k & -\Delta & 0\\
      0 & -\Delta^* & -\Sigma_k & 0\\
      \Delta^* & 0 & 0 & -\Sigma_k
  \end{array}\right).
\label{selfenergy}
\end{equation}
In here, the upgraded single particle excitation energy is now
$\Sigma_k=\epsilon_k-\mu+\langle T^{\rm Re}_k\rangle \,n$,
$\epsilon_k=\hbar^2 k^2/2m$ denotes the kinetic energy and
the gap energy is still defined as $\Delta=V^P p+\sum_i g_i\,\phi_i$.

\section{Thermodynamics}

In this paper we focus on the properties of thermodynamic equilibrium.
Thermodynamic equilibrium can be reached by demanding that the grand
potential $\Phi_G=-k_bT\ln\Xi$ at a fixed temperature has a minimal
value. In this definition $k_b$ is Boltzmann's constant, and $\Xi$
the partition function $\Xi={\rm Tr}[\exp(-\hat H_{\rm diag}/k_bT)]$.
The exponent containing the diagonalized Hamiltonian reads

\begin{figure}
\begin{center}\
  \epsfysize=60mm \epsfbox{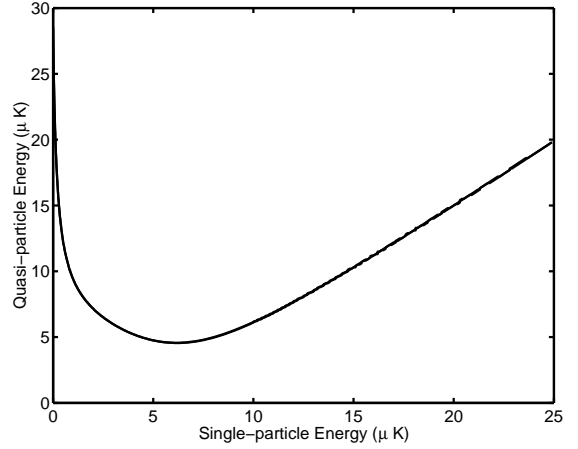}
\end{center}
\caption{Two overlapping quasi-particle energy spectra as a function of
  single-particle kinetic energy $E=\hbar^2k^2/(2m)$. The total
  density of the gas is $n=10^{14}$ cm$^{-3}$, the magnetic field is
  $B=900$ G, and the temperature is $T=0.01 T_F$. The energy-spectrum
  is calculated for two different values of the cutoff: i.e. $K=32k_F$
  and $K=64k_F$. The fact that the two lines are overlapping to the
  extent that the difference is difficult to see shows the
  renormalization in practice, demonstrating the validity of
  Eq.~(\ref{ren1})-(\ref{ren5}).} \label{fig7}
\end{figure}

\begin{eqnarray}
  \hat H_{\rm diag}&=&\sum_i (\nu_i-\mu_m)|\phi_i|^2\label{manybody}\\
  &&{}+\sum_{\smallvec{k}}\Bigl(\Sigma_k + E_k\bigl(
  \create{\alpha}{\smallvec{k}\uparrow}
  \annhilate{\alpha}{\smallvec{k}\uparrow}+
  \create{\alpha}{\smallvec{k}\downarrow}
  \annhilate{\alpha}{\smallvec{k}\downarrow}-1\bigr) \Bigr),\nonumber
\end{eqnarray}
which is a quadratic approximation to the original Hamiltonian.  The
energy spectrum $E_k$ results from a local diagonalization by the
Bogoliubov transformation of the self-energy matrix $\Sigma_{\boldvec k}$ at
each ${\boldvec k}$, where the obtained quasi-particle spectrum is
$E_k=\sqrt{\Sigma_k^2+\Delta^2}$. Note that the first summation term in
$\hat H_{\rm diag}$ results from a contribution from $\cal Q$ space,
and the second summation term from $\cal P$ space of
section~\ref{FBrep}.  The rotation to Bogoliubov quasi-particles
is given by the general canonical transformation

\begin{equation}
  \left(\begin{array}{c}
      \annhilate{\alpha}{\smallvec{k}\uparrow}\\
      \create{\alpha}{-\smallvec{k}\downarrow}
  \end{array}
\right)=\left(
\begin{array}{cc}
  \cos\theta & -e^{i\gamma}\sin\theta \\
  e^{i\gamma}\sin\theta & \cos\theta
\end{array}
\right)\left(
\begin{array}{c}
  \annhilate{a}{\smallvec{k}\uparrow}\\
  \create{a}{-\smallvec{k}\downarrow}
\end{array}
\right),
\end{equation}
where $\tan2\theta_k=|\Delta|/\Sigma_k$ is the Bogoliubov transformation
angle. The quasi-particle annihilation and creation operators are
indicated by $\annhilate{\alpha}{\smallvec{k}}$ and
$\create{\alpha}{\smallvec{k}}$. In Fig.~\ref{fig7} we show a typical
quasi-particle energy spectrum for $^6$Li versus the single-particle kinetic
energy, at a magnetic field of $B=900$ G and a temperature of $T=0.01
T_F$. The figure demonstrates how well the renormalizing
equations~(\ref{ren1})-(\ref{ren5}) work in obtaining a cutoff independent
energy spectrum. This is important because it implies that all the
thermodynamics which follow will also be $K$-independent.

For the stationary solution the grand potential, or equivalently, the free
energy, has indeed a minimum. This follows easily from setting the partial
derivative of the grand potential with respect to $\phi_i$ to zero: $\partial
\Phi_G/\partial \phi_i=0$. This gives the solution

\begin{equation}
  \phi_i=-\frac{\bar g_i p}{\bar \nu_i-\mu_m},
\end{equation}
which is also the stationary solution of Eq. (\ref{phi1}).
This equality is very useful because we can effectively eliminate the
molecular field from the equations. The quasi-particle states are now populated
according to the Fermi-Dirac distribution $n_k=[\exp(E_k/k_b T)+1]^{-1}$. The
mean fields are then determined by integrating the equilibrium single particle
density matrix elements, given by

\begin{eqnarray}
  n&=&\frac1{(2\pi)^2}\int_0^Kdk\, \bigl[(2n_k-1)\cos2\theta_k
  +1\bigr]\,,\\
  p&=&\frac1{(2\pi)^2}\int_0^Kdk\, (2n_k-1)\sin2\theta_k\,,
\label{eqfp}
\end{eqnarray}
Since $\theta_k$ depends on $n$ and $p$, these equations
require self-consistent solutions which are found from a numerical
iterative method.

In Fig. \ref{fig8} we show a plot of the chemical potential as a
function of temperature, for the case of $^6$Li in a homogenous gas,
at a magnetic field of $B=900$ G. Figure \ref{fig9} shows the ratio of
the critical temperature $T_c$ to the Fermi temperature $T_F$ as a
function of detuning. It clearly shows that there is a limiting value
of $T_c$ of about 0.5 $T_F$, similar to the value that has been
predicted for $^{40}$K in Ref.~\cite{holland}. The BCS result for the
critical temperature, given by the formula

\begin{equation}
  \frac{T_c}{T_F}\sim \exp{\left[-\frac{\pi}{2|a|k_F}\right]},
  \label{bcscrittemp}
\end{equation}
has been plotted in the same graph for comparison. The BCS line gives
a curve for $T_c$ higher than the resonance theory, since it does not contain
the energy-dependence of the $T$-matrix. The absolute value of the scattering
length in this magnetic field range is always larger than 2000 $a_0$, which
implies that $k_F|a|>1$---a clear indication that the BCS theory
breaks down in this regime.

\begin{figure}
\begin{center}\
  \epsfysize=60mm \epsfbox{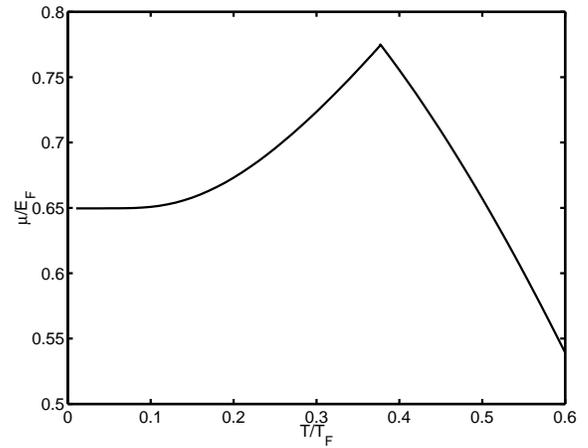}
\end{center}
\caption{Chemical potential as a function of temperature, for a magnetic field
  of $B=900$ Gauss for $^6$Li.} \label{fig8}
\end{figure}

\begin{figure}
\begin{center}\
  \epsfysize=60mm \epsfbox{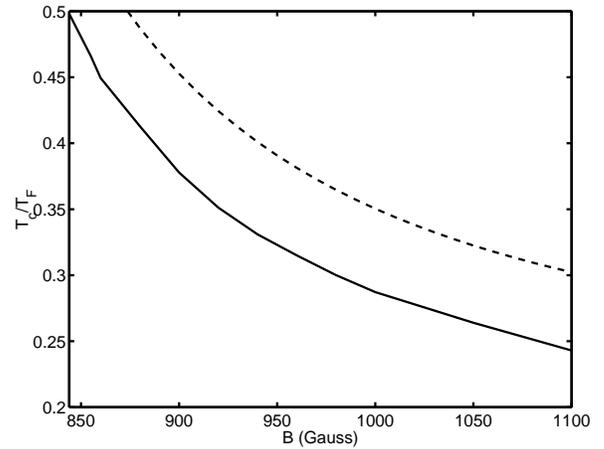}
\end{center}
\caption{Dependence of critical temperature on magnetic field for $^6$Li, for a
total   density of $n=10^{14}$ cm$^{-3}$ (solid line). The dashed line is,
  for comparison, the prediction of the regular BCS theory.}
\label{fig9}
\end{figure}

\begin{figure}
\begin{center}\
  \epsfysize=60mm \epsfbox{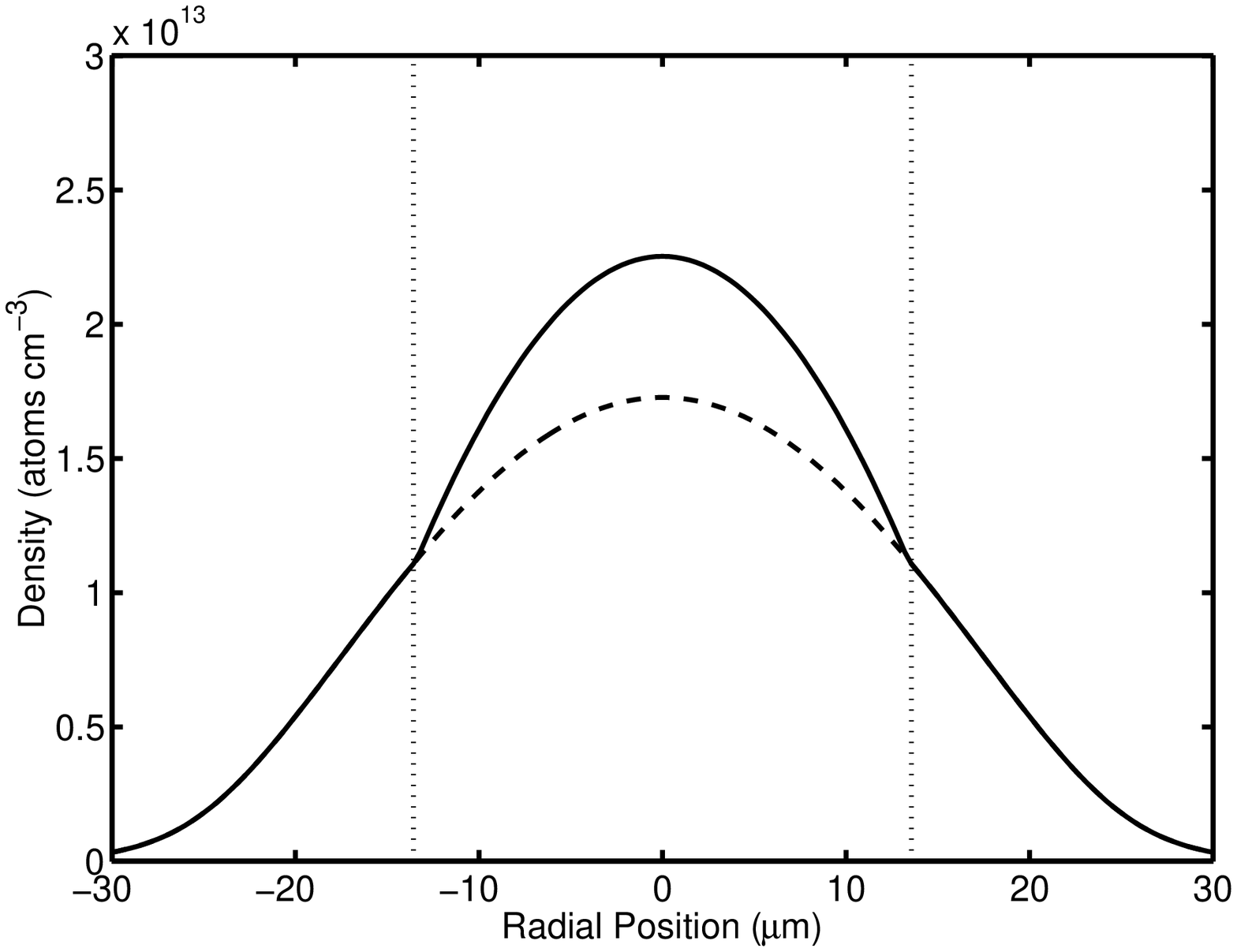}
\end{center}
\caption{
  Density profile for a gas of $^6$Li atoms (solid line), evenly distributed
among   the two lowest hyperfine states. The temperature is $T=0.2T_F$ at a
magnetic field of $B=900$ G. The trap-constant is $\omega=2\pi 500$ $s^{-1}$,
and we have a total number $N=5\times 10^5$ atoms. We compare this with a
profile resulting from the same $\mu$ (but for different total number $N$),
where artificially no superfluid is present by setting the pairing field $p$
equal to zero (dashed line).} \label{fig10}
\end{figure}

So far, our calculation has been done for a homogeneous gas. We will also
present results for a trapped lithium gas in a harmonic  oscillator potential
$V(\boldvec r)$ with a total number of  $N=5\times 10^5$ atoms, similar to what
we presented for $^{40}$K in  Ref.~\cite{chiofalo}. We treat the inhomogeneity
by making use of the  semiclassical local-density approximation, which involves
mainly the  replacement of the chemical potential by a spatially-dependent
version $\mu(\boldvec r)=\mu-V(\boldvec r)$. The thermodynamic  equations for
the homogeneous system are then solved at each point in  space~\cite{chiofalo}.
As a result, we obtain a spatially-dependent  density distribution. At zero
temperature, for a non-superfluid system, this gives the well-known Thomas-Fermi
solution. For a resonance system, however, a density  bulge appears in the
center of the trap, which is caused by a change  in compressibility when a
superfluid is present. This is shown in  Figure~\ref{fig10}, for a spherical
trap with a trap-constant of $\omega=2 \pi 500$ s$^{-1}$. This bulge is a
signature of  superfluidity and could experimentally be seen by fitting the
density  distribution in the outher wings to a non-resonant system, and thus
obtaining an excess density in the middle of the trap. For a discussion of the
abrupt change in the compressibility see Ref.~\ref{chiofalo}.

\section{Fluctuations in the mean fields and cross-over model}

In this section we make some comments on the connection between the resonance
superfluidity theory we have presented and related mean-field
approaches to discuss the cross-over of superconductivity from weak to
strong coupling. In the mean-field theory of BEC, most often reflected
in the literature by the Gross-Pitaevskii equation or finite
temperature derivatives, a small parameter is derived to justify the
application of the theory. This parameter,
$\sqrt{na^3}$~\cite{beliaev}, may be obtained from a study of higher
order corrections to the quasi-particle energy spectrum.  It has been
suggested that for a fermi system which exhibits superfluidity the
small parameter is given by a power of $k_F a$, and that the BCS
theory breaks down when this parameter approaches unity. However, the small
parameter in the theory of resonance superfluidity cannot be simply a function
of the scattering length for detunings close to resonance. This can
already be seen from the energy-dependence of the $T$-matrix, which
shows that around the Fermi-energy, the $T$-matrix may have an
absolute value much smaller than at zero energy where the scattering length is
defined. Moreover, even right on resonance when $\nu=0$ and the scattering
length passes through infinity, the $T$-matrix remains well behaved.

Instead of calculating the small parameter of this system, we choose a
different approach based on cross-over models between BCS and BEC,
formulated by Nozi\` eres \cite{nozieres}, and later expanded upon by
Randeria~\cite{randeria}. In the regular BCS theory for weakly coupled systems
the value of the critical temperature is given by the exponential dependence in
Eq.~(\ref{bcscrittemp}), but for strongly coupled systems this model results in
a logarithmically divergent prediction for $T_c$. The parameter $(k_F a)^{-1}$
is usually taken to describe the crossover from the weak coupling Bose
limit($(k_F a)^{-1} \rightarrow -\infty$) to the strong coupling ($(k_F
a)^{-1}\rightarrow +\infty$) BEC limit. The unphysical divergence in $T_c$
occurs because the process which dominates the transition in the weak coupling
regime is the dissociation of pairs of fermions. For a strongly coupled system,
however, the fermions are so tightly bound that the wave functions of pairs of
atoms begin to overlap, and the onset of coherence is signaled by excitations of
the condensed state, which occurs at a temperature well below the dissociation
temperature of the Cooper pairs. Thus, when moving from weak to strong coupling,
the nature of the transition changes from a BCS to a BEC type mechanism. An
explicit inclusion of the process of molecule formation, characterized by the
detuning, resonance width, and resonance position, will allow us to move from
one regime to the other.

The lowest order correction which connects between BCS and BEC type
superconductivity can be made by augmenting the density equation to
account for the formation of pairs of atoms. This is done by using the
thermodynamic number equation $N=-\frac{\partial \Phi_G}{\partial
  \mu}$, with $\Phi_G$ the total thermodynamic grand potential

\begin{equation}
  \Phi_G=\Phi_G^0-k_b T \Sigma_{{\mathbf q},i q_l}\ln \Gamma ({\mathbf
    q},iq_l).
\label{thermopot}
\end{equation}
The term $\Phi_G^0$ is a grand potential that does not include the
quasi-bound molecules and results from regular BCS theory. Retaining
only this term yields a theory which can only account for the free and
scattered fermionic atoms which contribute to the fermion density,
therefore the theory breaks down if a sizeable number of bound states
are formed. In the extreme limit of strong coupling, $\Phi_G^0$ becomes
negligible and equation (\ref{thermopot}) just reduces to the thermodynamic
potential of an ideal Bose gas. In this regime, the theory predicts the
formation of a condensate of molecules below the BEC transition temperature.

The function $\Gamma({\mathbf q},iq_l)$, which is a function of momentum
${\mathbf q}$ and thermal frequencies $i q_l$, is mostly negligible for a
weakly coupled system and has little affect on the value of $T_c$ in this
regime. It allows for the inclusion of the lowest contributing order of quantum
fluctuations~\cite{nozieres,randeria} by means of a general inclusion of
mechanisms for molecular pair-formation. In the resonance superfluidity model, a
similar term is present due to the formation of bosonic molecular bound states
$\phi_i$, and prevents the critical temperature from diverging. When the
coupling increases, the formation of molecules adds significantly to the total
density equation in both the cross-over models of superconductivity and in the
theory we have presented here. Moreover, the inclusion of the molecular term
allows for a smooth interpolation between the BCS and BEC limits. This is
clearly a substantial topic in its own right, and will be addressed further in a
future publication~\cite{milstein}.

\section{conclusions}

We have shown that it is possible to derive a mean-field theory of
resonance superfluidity, which can be applied to ultra-cold Fermi
gases such as $^6$Li and $^{40}$K. The Hamiltonian we use treats the
resonant states explicitly, and automatically builds the coupled scattering
equations into the many-body theory. With a study of analytical
scattering we have shown that these scattering equations can
completely reproduce a full coupled channels calculation for the
relevant energy regime. The energy-dependence of the $s$-wave phase
shifts can be described by a small set of parameters which correspond
to physical properties, such as the non-resonant background value of
the scattering length, and the widths and detunings of the Feshbach
resonances.  Close to resonance, we predict a large relative value of
0.5 $T_F$ for the critical temperature. The particular resonance under
study for $^6$Li occurs in the $(1/2,1/2)+(1/2,-1/2)$ collision
channel, and has its peak at $B_0=844$ G, and a width of about $\Delta B
\approx 185$ G~\cite{servaaspriv,abeelen}. This large width translates
into a large magnetic field range where the critical temperature is
within a factor of two from its peak value. This range is, for
comparison, much larger than for $^{40}$K. For $^6$Li there are also
two other Feshbach resonances, one in the $(1/2,-1/2)+(3/2,-3/2)$
state and another in the $(1/2,1/2)+(3/2,-3/2)$ state. They result
from coupling to the same singlet bound state, and occur at field
values of about $B_0=823$ G and $B_0=705$ G, and have a similar width
to the $(1/2,1/2)+(1/2,-1/2)$ resonance. The disadvantage of these
resonances, however, is that the atoms in these channels suffer from
dipolar losses, which are also resonantly enhanced. Three-body
interactions will be largely suppressed, as asymptotic p-wave
collisions will give very little contribution in the
temperature regime considered (an $s$-wave collision is always forbidden for at
least one of the pairs). From a study of cross-over models between BCS and BEC
we find no indication of break-down effects of the applied mean-field theory.

\section*{Acknowledgements}
We thank J. Cooper, E. Cornell, D. Jin, C. Wieman, B.J. Verhaar and B.
DeMarco for very stimulating discussions.  Support is acknowledged for
S.K.  and J.M. from the U.S. Department of Energy, Office of Basic
Energy Sciences via the Chemical Sciences, Geosciences and Biosciences
Division, and for M.C. from SNS, Pisa (Italy). Support is acknowledged for M.H.
and M.C. from the National Science Foundation, and for R.W. from the APART 
fellowship, Austrian Academy of Sciences.


\begin{references}
  
\bibitem{bec} M. H. Anderson, J. R. Ensher, M. R. Matthews, C. E.
  Wieman, and E. A. Cornell, Science {\bf 269}, 198 (1995); K. B.
  Davis, M.-O. Mewes, M. R. Andrews, N. J. van Druten, D. S. Durfee,
  D. M. Kurn, and W. Ketterle, Phys.\ Rev.\ Lett.\ {\bf 75}, 3969
  (1995); C. C. Bradley, C. A. Sackett, J.  J. Tollett, and R. G.
  Hulet, Phys.\ Rev.\ Lett.\ {\bf 75}, 1687 (1995); {\bf 79}, 1170(E)
  (1997).

\bibitem{vortices} J.E. Williams and M.J. Holland, Nature (London)
{\bf 401}, 568 (1999); M.R. Matthews, B. P. Anderson, P.C. Haljan, D.S. Hall,
C.E. Wieman, and E. A. Cornell M.R. Matthews {\it et al.}, Phys. Rev. Lett.
{\bf 83}, 2498 (1999); K.W. Madison, F. Chevy, W. Wohlleben, and J. Dalibard,
Phys. Rev. Lett.   {\bf 84}, 806 (2000), P.C. Haljan, I. Coddington, P. Engels,
and E.A. Cornell,   Phys. Rev. Lett. {\bf 87}, 210403 (2001), J.R. Abo-Shaeer,
C. Raman, J.M. Vogels, and W. Ketterle, Science {\bf 292}, 476 (2001).

\bibitem{dfg} B. DeMarco and D. S. Jin, Science {\bf 285}, 1703
  (1999).

\bibitem{truscott} G. Truscott, K. E. Strecker, W. I. McAlexander, G.
  B. Partridge, and R. G. Hulet, Science {\bf 291}, 2570 (2001).

\bibitem{schreck} F. Schreck, L. Khaykovich, K. L. Corwin, G.
  Ferrari, T. Bourdel, J. Cubizolles, and C. Salomon, Phys. Rev. Lett.
  {\bf 87}, 080403 (2001).

\bibitem{beliaev} S.T. Beliaev, Sov. Phys. JETP {\it 34}, 323 (1958).

\bibitem{abrikosov} A.A. Abrikosov, L.P. Gorkov, and I.E.
  Dzyaloshinski, {\it Methods of Quantum Field Theory in Statistical
    Physics}, Prentice-Hall Inc., New Jersey, (1963);

\bibitem{feshbach} H. Feshbach, Ann. Phys. {\bf 5}, 357 (1958); {\it
    ibid.}, Ann. Phys. {\bf 19}, 287 (1962), H. Feshbach, {\it
    Theoretical Nuclear Physics}, (Wiley, New York, 1992).

\bibitem{tiesinga} E. Tiesinga, B. J. Verhaar, and H. T. C. Stoof,
  Phys.  Rev. A {\bf 47}, 4114 (1993)

\bibitem{servaaspriv} S. Kokkelmans and B.J. Verhaar, (private
  communication).  The calculation is based on the analysis of the
  lithium interactions as described in \protect\cite{abeelen}. An
  alternative analysis has been described in \protect\cite{abraham}.

\bibitem{abeelen} F. A. van Abeelen, B. J. Verhaar, and A. J.
  Moerdijk, Phys.\ Rev.\ A {\bf 55}, 4377 (1997).

\bibitem{abraham} E. R. I. Abraham {\em et al.}, Phys.  Rev. A {\bf
    55}, R3299 (1997).

\bibitem{stoof} H.T.C. Stoof, J.M.V.A. Koelman, and B.J. Verhaar,
  Phys. Rev. B {\bf 38}, 4688 (1988).

\bibitem{bcs} J. Bardeen, L. N. Cooper, and J.~R.~Schrieffer, Phys.\
  Rev.  {\bf 108}, 1175 (1957); J. R. Schrieffer, {\it Theory of
    Superconductivity}, Perseus Books, Reading, Massachusetts, (1999).

\bibitem{leggett} A. G. Leggett, J. Phys. (Paris) {\bf C7}, 19 (1980);
  M.  Houbiers and H. T. C. Stoof, Phys.\ Rev.\ A {\bf 59}, 1556-1561
  (1999); G.  Bruun, Y. Castin, R. Dum {\em et al.}, Eur. Phys. J. D
  {\bf 7}, 433--439 (1999); H. Heiselberg, C. J. Pethick, H. Smith,
  and L. Viverit, Phys. Rev.  Lett. {\bf 85}, 2418 (2000).

\bibitem{taylor} J.R. Taylor, {\it Scattering Theory}, (Wiley, New
  York, 1972).

\bibitem{moerdijk} A.J. Moerdijk, B.J. Verhaar, and A. Axelsson, Phys.
  Rev. A {\bf 51}, 4852 (1995).

\bibitem{kokkelmans} S.J.J.M.F. Kokkelmans, B.J. Verhaar, and K.
  Gibble, Phys.  Rev. Lett. {\bf 81}, 951 (1998).

\bibitem{leo} P.J Leo, E. Tiesinga, P.S. Julienne, Phys. Rev. Lett.
  {\bf 81}, 1389 (1998)

\bibitem{kempen} E.G.M. van Kempen, S.J.J.M.F. Kokkelmans, D.J.
  Heinzen, and B.J. Verhaar, to be published in Phys. Rev. Lett.

\bibitem{houbiers} M. Houbiers, H.T.C. Stoof, W.I. McAlexander, and R.G.
Hulet, Phys. Rev. A {\bf 57}, R1497 (1998).

\bibitem{vogels} J.M. Vogels, B.J. Verhaar, and R.H. Blok, Phys. Rev.
  A {\bf 57}, 4049 (1998).

\bibitem{holland} M. Holland, S.J.J.M.F. Kokkelmans, M.L.
  Chiofalo, and R. Walser, Phys. Rev. Lett. {\bf 87}, 120406 (2001).

\bibitem{chiofalo} M.L. Chiofalo, S.J.J.M.F. Kokkelmans, J.N.
  Milstein, and M.  Holland, submitted to Phys. Rev. Lett.

\bibitem{peletminskii} A. I. Akhiezer and S. V. Peletminskii, {\it
    Methods of Statistical Physics}, Pergamon Press Ltd., Oxford,
  England (1981).

\bibitem{walser599} R.~Walser, J.~Williams, J.~Cooper, M.~Holland,
  Phys. Rev. A. {\bf 59}, 3878 (1999).

\bibitem{walser999} R. Walser and J. Cooper and M. Holland, Phys.
  Rev. A {\bf 63}, 013607 (2001).

\bibitem{timmermans} E. Timmermans {\it et al.}, Phys. Rev. Lett. {\bf
    83}, 2691 (1999).

\bibitem{abeelen2} F.A. van Abeelen and B.J. Verhaar, Phys. Rev. Lett.
  {\bf 83}, 1550 (1999).

\bibitem{holland2} M. Holland, J. Park, and R. Walser, Phys. Rev.
  Lett. {\bf 86}, 1915 (2001)

\bibitem{kokkelmans2} S.J.J.M.F. Kokkelmans, H. M. J. Vissers, and B. J.
Verhaar, Phys. Rev. A {\bf 63}, 031601 (2001).

\bibitem{blaizot} J. P. Blaizot and G. Ripka, {\it Quantum Theory of
      Finite Systems}, The MIT Press, Cambridge, Massachusetts (1986).

\bibitem{nozieres} P. Nozi\` eres and S. Schmitt-Rink, J. Low Temp.
  Phys. {\bf 59}, 195 (1982).

\bibitem{randeria} See M. Randeria in {\it Bose-Einstein
    condensation}, ed. by A.  Griffin, D.W. Snoke and S.  Stringari,
  Cambridge Un. Press, Cambridge (1995).

\bibitem{milstein} J.N. Milstein, S.J.J.M.F. Kokkelmans, R. Walser,
  and M.J.  Holland, to be published.

\end{references}
\end{document}